\documentclass[twocolumn]{aastex63}

\received{XXX}
\revised{XXX}
\accepted{XXX}
\submitjournal{Planetary Science Journal}

\shorttitle{ALMA Observations of Io in/out of Eclipse}

\begin{document}

\title{ALMA Observations of Io Going into and Coming out of Eclipse}

\author[0000-0002-4278-3168]{Imke de Pater}
\affiliation{University of California, 501 Campbell Hall, Berkeley, CA 94720, USA,
and Faculty of Aerospace Engineering, Delft University of Technology, Delft 2629 HS, The Netherlands}

\author[0000-0001-9867-9119]{Statia Luszcz-Cook}
\affiliation{University of Columbia, Astronomy Department, New York, USA}

\author[0000-0002-1607-6443]{Patricio Rojo}
\affiliation{Universidad de Chile, Departamento de Astronomia, Casilla  36-D, Santiago, Chile}

\author{Erin Redwing}
\affiliation{University of California, 307 McCone Hall, Berkeley, CA 94720, USA}

\author[0000-0002-9068-3428]{Katherine de Kleer}
\affiliation{California Institute of Technology, 1200 East California Boulevard, Pasadena, CA 91101, USA}

\author[0000-0002-9820-1032]{Arielle Moullet}
\affiliation{SOFIA/USRA, NASA Ames Building N232, Moffett Field, CA 94035, USA}

\vskip4.0truein



\pagebreak

\begin{abstract}

  We present 1-mm observations constructed from ALMA [Atacama Large
  (sub)Millimeter Array] data of SO$_2$, SO and KCl when Io went from
  sunlight into eclipse (20 March 2018), and vice versa (2 and 11
  September 2018).  There is clear evidence of volcanic plumes on 20
  March and 2 September. The plumes distort the line profiles, causing high-velocity ($\gtrsim$500 m/s)
  wings, and red/blue-shifted shoulders in the line profiles.  During
  eclipse ingress, the SO$_2$ flux density dropped exponentially, and the atmosphere reformed in a linear
  fashion when re-emerging in sunlight, with a ``post-eclipse brightening'' after $\sim$10 minutes. While both the in-eclipse decrease
  and in-sunlight increase in SO was more gradual than for SO$_2$, the
  fact that SO decreased at all is evidence that self-reactions at the surface are important and fast, and that
  in-sunlight photolysis of SO$_2$ is the dominant source of SO.
  Disk-integrated SO$_2$ in-sunlight flux densities are $\sim$2--3 times higher than in-eclipse, indicative of a roughly 30--50\%
  contribution from volcanic sources to the atmosphere. Typical column
  densities and temperatures are  $N \approx (1.5 \pm 0.3) \times 10^{16}$ cm$^{-2}$ and
  $T \approx 220-320$ K both in-sunlight and in-eclipse, while the fractional coverage of the gas
   is 2--3 times lower in-eclipse than in-sunlight. The low level
   SO$_2$ emissions present during eclipse may be sourced by stealth
   volcanism or be evidence of a layer of non-condensible gases
   preventing complete collapse of the SO$_2$ atmosphere. The melt in
   magma chambers at different volcanoes must differ in composition to
   explain the absence of SO and SO$_2$, but simultaneous presence of KCl over Ulgen Patera.

\end{abstract}



\pagebreak


\section{Introduction} \label{sec1}

Jupiter's satellite Io is unique amongst bodies in our Solar System.
Its yellow-white-orange-red coloration is produced by SO$_2$-frost on
its surface, a variety of sulfur allotropes (S$_2$--S$_{20}$), and
metastable polymorphs of elemental sulfur mixed in with other species
(Moses and Nash, 1991). Spectra of the numerous dark calderas, sites of intermittant
volcanic activity, indicate the presence of (ultra)mafic minerals such
as olivine and pyroxene (Geissler et al., 1999). When Io is in eclipse
(Jupiter's shadow), or during an ionian
night (visible only from spacecraft), visible and near-infrared images of the satellite reveal dozens of
thermally-bright volcanic hot spots (e.g., Geissler et al., 2001; Macintosh et al., 2003; de Pater
et al, 2004; Spencer et al., 2007; Retherford
et al., 2007). This widespread volcanic activity is
powered by strong tidal heating induced by Io's orbital eccentricity,
which is the result of the Laplace orbital resonance between Io,
Europa, and Ganymede. Some volcanoes are associated with active
plumes, which are a major source of material into Io's atmosphere,
Jupiter's magnetosphere, and even the interplanetary medium. The
mass-loss from Io's atmosphere is estimated at 1 ton/second (Spencer and Schneider, 1996), yet
the atmosphere is consistently present, indicating an ongoing
replenishment mechanism. However, the amount of material pumped into
Io's atmosphere by volcanism is not well known, and it is consequently
not known whether the dynamics in Io's atmosphere is primarily driven
by sublimation of SO$_2$ frost on its surface or by
volcanoes. An additional source of atmospheric gas may be sputtering from Io's surface.

\added{A decade after the inital detection of gaseous SO$_2$ in its
  $\nu_3$ band (7.3 $\mu$m) from
the Voyager data (Pearl et al., 1979),  Io's ``global''
SO$_2$ atmosphere was detected at 222 GHz  (Lellouch et al.,
1990). These data } \deleted{Io's
``global'' SO$_2$ atmosphere was first detected at 222 GHz, and} 
revealed a surface pressure of order 4--40 nbars
(2$\times 10^{17}$--2$\times 10^{18}$ cm$^{-2}$), covering 3--20\% of
the surface, at temperatures of $\sim$~500-600 K. There is a large uncertainty in the temperature, however,
since it is extremely difficult to disentangle the contributions of
density, temperature and fractional coverage\footnote{The fractional coverage
    of the gas is the fraction of the projected surface that
    is covered by the gas.} in the line profiles
(e.g., Lellouch et al., 1992). Moreover, \added{zonal winds would
  broaden the line profile (``competing'' with temperature), while
  Ballester et al (1994) noted that winds from volcanic eruptions may
  distort the lineshape,} \deleted{velocities, such as expected
from volcanic eruptions, may also effect the lineshape (Ballester et
al., 1994),} \added{both} adding complications to modeling efforts. Although SO$_2$ 
has now been observed at mm, UV, and at thermal infrared
wavelengths, its temperature and column density are
still poorly constrained.

\added{Based upon photochemical considerations alone, in a
  SO$_2$-dominated atmosphere one would expect at least the products
  SO, O$_2$, as well as atomic S and O (e.g., Kumar, 1985; Summers, 1985). SO was detected at
  microwave wavelengths at a
  level of a few \% compared to the SO$_2$ abundance (Lellouch,
  1996). While O$_2$ has not (yet) been detected, S and O have been
  detected e.g., in the form of auroral emissions off Io's limb along
  its equator (e.g., Geissler et al. 2004).
 We further note that gaseous NaCl was first detected by Lellouch et al. (2003), and a
  tentative detection of KCl was reported by Moullet et al. (2013). Both NaCl and KCl were mapped with ALMA by
Moullet et al. (2015). }

Spatially resolved data obtained with the Hubble Space Telescope (HST)
at UV wavelengths revealed that SO$_2$ was mainly confined to
latitudes within 30--40$^\circ$ from the equator, with a higher column
density and latitudinal extent on the anti-jovian side \added{(central meridian
longitude CML $\sim$180$^\circ$W)} (e.g., Roesler
et al., 1999; Feaga et al., 2004, 2009). The sub- (\deleted{central meridian
longitude}CML $\sim$0$^\circ$W) to anti-jovian \deleted{(CML$\sim$ 180$^\circ$W)}
hemisphere distribution was
confirmed using disk-averaged thermal infrared data of the 19-$\mu$m $\nu_2$ band of
SO$_2$, observed in absorption against Io with the TEXES instrument on
NASA's Infrared
Telescope Facility (IRTF) in 2001--2005 (Spencer et al., 2005). While these observations 
showed a temperature of $\sim$115-120 K, interpretation of
disk-resolved observations of the SO$_2$ $\nu_1 +
\nu_3$ band at $4\;\mu$m with the CRIRES instrument on the Very Large Telescope (VLT) favors a 
temperature of $\sim$170 K (Lellouch et al.,
2015). Typical column densities in all these data vary roughly from
$\sim 10^{16}$ on the sub-jovian hemisphere 
to $\sim 10^{17}$ cm$^{-2}$ on the anti-jovian side.

Moullet et al. (2010) used the spatial distribution derived from the
HST/UV and TEXES/IRTF observations to analyze SO$_2$ maps at mm-wavelengths obtained with 
the Sub-Millimeter Array (SMA). By decreasing the number of free
parameters to just temperature and column density, using the
fractional coverage from the UV and mid-IR data, they 
derived a disk-averaged column density of
2.3--4.6$\times 10^{16}$ cm$^{-2}$ and temperature between 150--210 K
on the leading (CML $\sim$ 90$^\circ$W)
hemisphere, and 0.7--1.1$\times 10^{16}$ cm$^{-2}$ with 215--255 K on
the trailing (CML $\sim$ 270$^\circ$W) side. These temperatures and
column densities are considerably lower than the earlier mm-wavelength measurements.

As mentioned above, it is still being debated whether the primary
source of Io's atmosphere is volcanic or driven by sublimation,
although it is clear that both volcanoes and SO$_2$ frost do play a
role (Jessup et al., 2007; Lellouch et al., 1990, 2003, 2015; Moullet
et al., 2010, 2013, 2015; Spencer et al., 2005; Tsang et al.,
\added{2012,} 2016). Although much of the
SO$_2$ frost may ultimately have been produced by volcanoes, the extent
to which volcanoes directly affect the atmosphere is unknown;
moreover, this likely varies over time. \added{Mid-IR observations
  showed an increase in the SO$_2$ abundance with decreasing
  heliocentric distance, which is, at least in part, in support of the
  sublimation theory (Tsang et al., 2012). Further support was given
  by the analysis} of the SMA maps mentioned
above\added{, which indicated} \deleted{suggests} that frost sublimation is the
main source of gaseous SO$_2$, and photolysis of SO$_2$ is the main source of
SO, since volcanic activity is not sufficient to explain the SO column
density and distribution (Moullet et al., 2010).
On the other hand, SO$_2$ gas is
enhanced above some volcanic hot spots ( McGrath et al., 2000\deleted{; Pearl
et al., 1979}) , and Pele's plume contains the sulfur-rich gases S$_2$, S,
and SO (\added{McGrath et al., 2000,} Spencer et al., \replaced{2005}{2000}; Jessup et al., 2007), indicative of
volcanic contributions to Io's atmosphere. 
\added{For more information on the pros and cons of the driving forces (sublimation vs
  volcanic) of Io's atmospheric dynamics, see the} excellent reviews of the state-of-knowledge of Io's atmosphere in the
mid-2000's by McGrath et al. (2004) and
Lellouch et al. (2007).

Observations of Io right before, after and during an eclipse provide the best way to separate the volcanic
from sublimation-driven contributions to its atmosphere.  The
atmospheric temperature is expected to drop within minutes after Io
enters an eclipse (e.g., de Pater et al., 2002).  The SO$_2$ gas that makes
up the bulk of Io's atmosphere is expected to condense out on a
similar timescale, set by the vapor pressure of this gas, which is a
steep exponential function of temperature
($P_{vapor} = 1.52 \times 10^8 e^{-4510/T}$ bar; Wagman, 1979).

Tsang et al (2016) obtained the first direct obervations of the 
SO$_2$ $\nu_2$ band in
Io's atmosphere in eclipse with the TEXES instrument on the Gemini
telescope. Their disk-integrated spectra were sensitive to surface
temperature, atmospheric temperature, and SO$_2$ column
abundance. Based on a simple model with a surface temperature of 127
K, they  
found that this value dropped to 105 K within minutes 
after entering eclipse. A range of models for Io's
atmospheric cooling all showed that the SO$_2$ column density
simultaneously dropped, by a factor of 5$\pm$2. They therefore
concluded that the atmosphere must contain a large component that is driven by
sublimation.

Although the radical SO will not condense at these temperatures, it
may be rapidly removed from the atmosphere through reactions with Io's surface
(Lellouch et al. 1996). However, a bright emission band complex at
1.707 $\mu$m, the forbidden electronic
$a^1 \Delta \rightarrow X^3 \Sigma^-$ transition of SO,
was observed in a disk-integrated spectrum of Io while in eclipse.
Based on the linewidth, a rotational temperature of $\sim$~1000 K was
derived, and the authors concluded that excited SO molecules were
ejected from the then very active volcano Loki \added{Patera} (de Pater et al.,
2002).

More recent observations reveal the spatial
distribution of SO, and show that the correlation with volcanoes is
tenuous at best (de Pater et al., 2007; 2020). 
Both the spatial distribution and the spectral shape of the SO
emission band vary considerably across Io and over time. In their most
recent paper (de Pater et al., 2020), the authors
suggest that the emissions are likely caused by a large number of
stealth plumes, ``high-entropy'' eruptions (Johnson et al., 1995) produced through the interaction of silicate melts
with superheated SO$_2$ vapor at depth. These plumes do not have much
dust or condensates, and are therefore not seen in reflected
sunlight. The SO data are further suggestive
of non-LTE processes, in addition to the direct ejection of excited SO
from the volcanic vents.

In order to shed more light on the core question whether the dynamics
in Io's atmosphere is predominantly driven by sublimation of SO$_2$-ice or volcanic activity, we 
present spatially resolved observations of the
satellite at 880 $\mu$m 
when Io moved from sunlight into eclipse, and half a year later from
eclipse into sunlight. The observations and data reduction are
discussed in Section 2 with
results presented in Section 3. The analysis of line profiles is
presented in Section 4, with a
discussion in Section 5.  Conclusions are summarized in Section 6.

\section{Observations and Data Reduction}\label{sec2}

We observed Io with the Atacama Large (sub)Millimeter Array (ALMA) on
20 March 2018 just before and after the satellite moved into eclipse.
Similar experiments were conducted when Io moved
out of eclipse on 2 and 11 September, 2018. Figure~\ref{fig:geometry}
shows the viewing geometry on both occasions. All observations were
conducted in Band 7, the 1-mm band. Each continuous observation of a
source (calibrator
or Io) is referred to as a scan, and gets a scan ``label''.
Amplitude and bandpass calibrations were performed on the radio source
J1517-2422 during the first $\sim$15 min of each of the six $\sim$35-min
long observing sessions (two sessions on each date). 
The phases were calibrated on J1532-1319 in March and on
J1507-1652 in September. These observations or scans (typically 30--60 sec
long) were taken before,
interspersed, and at the end of the Io observations. Typical Io scans
are 6--7 min \added{long}, though towards the end of the 
observing sessions they usually lasted for only 1--2 minutes. 
The flux densities of J1517-2422 were checked
with the ALMA calibrator catalogue; no updates were needed since the
observatory's initial pipeline data reduction. On 11 September the flux densities
of both Io and the secondary calibrator were much lower for the
in-eclipse data than for the ones in-sunlight, perhaps caused by some
decorrelation in the phases\added{ and/or pointing errors}. We therefore multiplied the Io-in-eclipse data
by a factor of 1.15, the ratio for the secondary calibrator between
the in-sunlight and in-eclipse data sets.

We observed several transitions of
SO$_2$ and SO, and one transition of KCl. These transitions, together
with the spectral window (spw) used to observe them, the total
bandwidth and channel width of each spw are listed in
Table 1. We typically had 3--4 beams (resolution elements)
across the satellite. Usually all scans on a particular source are
combined to create a map or a spectral line data-cube. Since we are interested in particular in how the
spatial brightness distribution and flux density change during 
eclipse ingress and egress, we imaged individual scans, and even
fractions of a scan, as summarized in Table 2. 

After the calibration and initial flagging was done in the ALMA pipeline,
we split off the Io data into its own dataset (referred to as a
measurement set, Io.ms) and attached a new ephemeris file so that
the position and velocity got updated every minute of time. We used
the Common Astronomy Software Applications package, CASA, version 5.4.0-68 for
all our data reduction. This version properly handles the tracking of
Io's motion across the sky and velocity along the
line-of-sight. Our final products are centered on Io, both in space
(images) and velocity (line profiles).
We first created continuum maps of the satellite, used initially for additonal
flagging and selfcalibration of the data (e.g., Cornwell and Fomalont 1999). Mapping was done using
tCLEAN; a model of Io's continuum emission served as a
``startmodel'' in the deconvolution (``cleaning'') and
selfcalibration process. The model is a uniform
limb-darkened disk with a disk-averaged brightness temperature that
matches the data (typically between 65 and 80 K), and a limb-darkening
coefficient $q=$0.3 (i.e., the brightness falls off towards the limb as
cos$\theta^q$, with $\theta$ the emission angle). All data were
selfcalibrated twice (phase selfcal only), although the second selfcalibration did
not improve the data substantially over the first one. 

Before creating spectral line data-cubes, we split out each spectral
window (Table 1) into its own measurement set (Io-spwx.ms, with x=0--7), and subtracted the
continuum emission from each Io-spwx.ms (x=1--7) dataset using the CASA routine UVCONTSUB. At this
point we have spectral image data-cubes of just the emission of each species
(SO$_2$, SO, KCl).

In order to create maps of the brightness distribution of each species at a high
signal-to-noise (SNR), we averaged the data in velocity over 0.4 km/s, centered at
the center of each line; these maps are referred to as ``line center maps''. To evaluate line profiles, we also
constructed 3-dimensional (3D) data-cubes with RA and DEC along the x-
and y-axes, and frequency (or velocity) along the z-axis, where each image
plane was averaged over 0.142 km/s, which translates roughly into a
frequency resolution of $\sim$160 kHz\footnote{Note that the precise number depends on
frequency $\nu$, since $\Delta \nu = \nu v/c$, with $v$ the velocity
and $c$ the speed of light.}, slightly larger than the 122 kHz width
of an individual channel in each spw. 
All (spectral line, line center, and continuum) maps were constructed using uniform
weighting, and cleaned using the Clark or H\"ogbom algorithm \added{with a
gain of 10\% in CASA's tCLEAN routine. In essence, in this routine we
iteratively remove 10\% of the peak flux density from that location on the map, together with the
synthesized beam (the telescope's antenna
pattern). This process is repeated until essentially only noise is
left  in the ``residual'' map. These socalled 
``clean components'' form a map, the .model map in CASA. An example is
  shown in Figure~\ref{fig:cc}. (Note that the
continuum maps were deconvolved using a startmodel in tCLEAN, as described
above). The .model map is convolved with 
a circular gaussian beam with a full width at
half power (HPBW) that best matches the inner part of the synthesized
beam (see Table 2 for the HPBW values) before being added back to the
residual map. The .model map in Fig.~\ref{fig:cc} shows the clean
components of the map displayed in the top left panel in
Figure~\ref{fig:MarchSO2}, discussed below in Section \ref{sec3a}.} We used a cell (or pixel) size
for all maps of 0.04", i.e., between 5.5--9 pixels/beam.

\section{Results}\label{sec3}

\subsection{Continuum Maps}\label{sec3.1}

The continuum maps for each of the 6 sessions, three in-sunlight and
three in-eclipse, are
very similar, and do not show any structure
other than that the maximum temperature is not centered on Io, but
slightly displaced towards the afternoon, as shown in
Figure~\ref{fig:cont}a,b. \deleted{Note, though, that the beam size is large
(0.22--0.35''; Table 2) compared to typical variations in albedo.}
We determined the total flux density from such
maps, since it is impossible to determine this from the {\it uv}-data, as
these are dominated by the signal from nearby Jupiter. For the maps
in-eclipse, the sidelobe patterns from Jupiter produce broad (similar
size as Io) low-level
(few \% of Io's peak intensity) negative and/or
positive ripples which affect the precise determination of Io's
flux density. Although ideally one would subtract Jupiter from
the visibility data, in practice this is not easy as Jupiter is not a
uniform disk at mm-wavelengths (e.g., de Pater et al., 2019), moves with respect to Io,
is mostly resolved out, and mostly on the edge or outside the
$\sim$20'' primary beam.
We therefore opted to correct for these negative or positive backgrounds
by subtracting the average flux density per pixel as determined from
an annulus around Io in each of the 6 maps. Each map was constructed
from all scans in the particular observing session, although the
in-eclipse scan 6 on 11 September was not used (too much affected by nearby Jupiter).

The total flux density, $F_J$, normalized to a geocentric distance of
5.044 AU (Io's diameter is 1'' at this distance) and averaged over all
six measurements, $F_J =$ 5.43$\pm$0.15 Jy, which translates into a
disk-averaged brightness temperature, $T_b = $ 93.6$\pm$2.5 K.  Since
Io blocks the cosmic microwave background radiation (CMB), which is
0.044 K at this wavelength, we have added this value to all brightness
temperatures quoted.  The uncertainties quoted above are the standard
deviation or rms spread in the continuum measurements, which is much
larger than the uncertainty in any single continuum measurement based
on the rms in the maps, which varies from 0.001--0.008 Jy (0.02--0.14
K) in-sunlight, to 0.008--0.05 Jy (0.14--0.86 K) in-eclipse. For
absolute values we need to add the calibration uncertainty in
quadrature. A typical calibration error for ALMA data is $\sim$5\%,
i.e., the total uncertainty on the brightness temperature is $\sim$5
K.

For all three dates, there is a small
difference between $T_b$ in sunlight and in eclipse: in sunlight we
find $F =$ 5.43$\pm$0.11 Jy (5.43, 5.39, 5.59 Jy, in chronological order), i.e., $T_b =$ 93.6$\pm$1.8 K, and in
eclipse $F=$ 5.25$\pm$0.14 Jy (5.16, 5.25, 5.44 Jy), i.e., $T_b =$ 90.8$\pm$2.2 K.  The
uncertainties are again the rms spread in the data points. The
difference between these numbers, and because all in-eclipse values are
lower than the corresponding in-sunlight values, suggests that $T_b$
may decrease by $\sim$3 K after entering eclipse. \replaced {This is
interesting, as Tsang et al. (2016) saw a drop in surface temperature
at 19 $\mu$m from 127 K to 105 K within minutes after entering
eclipse.} {This is
  interesting since at mid-IR wavelengths Io's surface
  temperature dropped steeply within minutes after entering eclipse
  (Morrison and Cruikshank, 1973; Sinton and Kaminsky, 1988). Tsang et
  al. (2016)  measured a drop in surface temperature at 19 $\mu$m from
  127 K to 105 K.} At radio wavelengths we typically probe $\sim$10--20 wavelengths
deep into the crust, or $\sim$1--2 cm at the ALMA wavelengths
used \added{(for pure ice this can be hundreds of wavelengths deep)}. Hence, even after having been in shadow for $\sim$2 hours, the
temperature at depth had decreased by no more than $\sim$3 K.

The morning-afternoon asymmetry in Io's continuum brightness is also
still present during eclipse, or after having been in darkness for $\sim$2 hrs (Fig.~\ref{fig:cont}b), which further supports our finding that eclipse cooling at depth is slow. 
\deleted{We note, though, that it may be possible that
part of the asymmetric brightness distribution in-eclipse is affected by thermal emission from Loki
Patera; our data have too low a spatial resolution and too high a
noise level to distinguish any potential effect of Loki Patera from
the morning--afternoon asymmetry.}
We used a simple thermal conduction model (after the infrared version
of the de Kleer et al. 2020 model), neglecting albedo variations
across Io's surface (which \replaced{is}{may be} reasonable given our
relatively low spatial resolution), to demonstrate that the difference
between the eclipse cooling at infrared and mm wavelengths can be
explained if the upper $\sim$cm or so of Io's crust is composed of two
layers.  For the model in Figure 2c we assumed \added{a bolometric Bond
  albedo, $A=$0.5}, an infrared emissivity \added{$\epsilon =$}0.9 and
a thermal inertia \added{$\Gamma =$}50 J m$^{-2}$ K$^{-1}$
s$^{-1/2}$. \added{This number is similar to the value of 70 J
  m$^{-2}$ K$^{-1}$ s$^{-1/2}$ derived by Rathbun et al. (2004) from
  {\it Galileo}/PPR data.} At millimeter wavelengths we assumed
\added{$A=$0.5, $\epsilon=$0.78 and $\Gamma =$} \deleted{emissivity of 0.78
  and a thermal inertia of}320 J m$^{-2}$ K$^{-1}$ s$^{-1/2}$. These
models, for a ``typical'' surface location at mid-latitudes, more or
less match the data, and suggest that Io's surface is overlain with a
thin (no more than a few mm thick) low-thermal-inertia layer, such as
expected for dust or fluffy deposits from volcanic plumes, overlying a
more compact high-thermal-inertia layer, composed of ice (likely
coarse-grained, and/or sintered) and rock. This is very similar to the
model proposed by Morrison and Cruikshank (1973) based upon seven
eclipse ingress or egress measurements at a wavelength of 20 $\mu$m,
although our value for the low thermal inertia layer is $\sim$4 times
higher. \added{Sinton and Kaminsky (1988) analyzed 13 observations of
  eclipse ingress and egress in the early 1980s at wavelengths between
  3.5 and 30 $\mu$m. They found a best fit by assuming Io to be
  covered by both dark ($A=$0.10) and bright ($A=$0.47) areas, with
  $\Gamma =$5.6 and 50  J m$^{-2}$ K$^{-1}$ s$^{-1/2}$, resp., where the low thermal inertia layer is
  just a thin layer atop a much higher thermal inertia. They noted
  that cooling was rapid during the first few minutes, followed by a
  slower process which they attributed to a combination of the higher thermal inertia,
  higher albedo passive component and emission from hot
  spots. Although their thermal inertias are roughly an order of
  magnitude smaller than the values we found, the overall physical
  picture of a thin dusty/porous layer atop a more compact high
  inertia layer is the same for all models. Our millimeter data in
  particular add a
  strong constraint to the higher thermal inertia layer roughly a cm
  or so
  below Io’s surface, a depth not probed at shorter wavelengths.}  Our values for the upper dusty layer are
\added{also} similar to those reported for the other galilean
satellites (e.g., Spencer, 1987; Spencer et al., 1999; de Kleer et
al., 2020), \added{ and they agree with the best-fit values found in the
  thermophysical parametric study by Walker et al. (2012). Note,
  though, that the latter study, as well as other 2-component thermal
  inertia studies at mid-IR wavelengths, all refer to horizontal surface
  variations, while our study refers to a vertically stacked model. In
  a future paper we intend to expand our 2-layer model to include proper dark
  and bright surface areas, as done for Ganymede in de Kleer et al. (2020).}

\subsection{Line Center Maps (averaged over 0.4 km/s)}\label{sec3a}

\vskip 0.15truein

\subsubsection{Line center Maps on 20 March 2018}\label{sec3a.1}

\paragraph{SO$_2$ maps:}

Figure~\ref{fig:MarchSO2} (top row) shows SO$_2$ maps at 346.652 GHz (spw2),
averaged in velocity over 0.4 km/s ($\sim$0.45 MHz) centered on the line, from
20 March 2018 when Io went into eclipse. The bottom row shows
simultaneously taken SO maps
(averaged over both transitions to increase the SNR). The first panel shows Io-in-sunlight,
and the next 2 panels show the satellite $\sim$~6 min
\deleted{(scan 7a)} and $\sim$~15 min \deleted{(scans 11$+$15)} after entering eclipse
(Table 2). The large circle shows the outline of Io, as determined
from simultaneously obtained images of the continuum emission. 
As soon as Io enters an eclipse, the atmospheric and surface
temperatures drop \added{(Fig.~\ref{fig:cont})}, and SO$_2$ is expected to condense out on a time scale t $\sim$
H/c$_s \approx$ 70 sec, for a scale height H $\approx$ 10 km and
sound speed c$_s \approx 1.5 \times 10^4$ cm/s (de Pater et al.,
2002)\added{, unless a layer of non-condensible gases prevents
complete collapse (Moore et al., 2009)}. Once
in eclipse, \added{assuming complete collapse,} the only SO$_2$ we see should be volcanically sourced. The
letters on Figure~\ref{fig:MarchSO2} show the positions of Karei Patera (K), Daedalus Patera (D),
and North Lerna (L). Due to the excellent match between the location
of these volcanoes and the SO$_2$ emissions on this day, these
volcanoes are likely the main sources of
SO$_2$ gas for Io-in-eclipse. All three volcanoes have shown
either plumes or changes on the surface attributed to plume activity in the
past (Geissler et al., 2004; Spencer et
al., 2007).

\paragraph{SO maps:}

 \added{ SO can be volcanically sourced, i.e., produced in
  thermochemical equilibrium in the vent (Zolotov and Fegley, 1998),
  or later via the reaction O $+$ S$_2$ at a column-integrated rate of
  $4.6\times10^{11}$ cm$^{-2}$ s$^{-1}$, or while in sunlight it can
  be produced through photolysis of SO$_2$ at a similar
  column-integrated rate (Moses et al., 2002). About 70\% of SO is
  lost through photolysis into S and O, but during an eclipse the only known
  loss is through a reaction with itself: 2SO $\rightarrow$ SO$_2$ $+$
  S, at a rate of $3.25\times10^{10}$ cm$^{-2}$ s$^{-1}$ (Moses et
  al., 2002). Hence, to eliminate an entire column of 10$^{15}$
  cm$^{-2}$ (Section 4.3) would take 8.5 hrs, or almost an hour to
  loose a 10$\times$ smaller column. Hence one would not expect much
  change in the SO column density upon eclipse ingress. The data,
  however, clearly show a decrease in the SO emission}
\deleted{Although SO is much more volatile than SO$_2$, it may still
  be rapidly removed from the atmosphere because SO is highly reactive
  with itself on the surface (Lellouch, 1996). Indeed, we do see the
  SO emission decreasing} after eclipse ingress, though not as fast as
for SO$_2$. \added{The observed decrease suggests that SO may be much
  more reactive with itself than captured by the above
  reaction rate. Additional (in-between) reactions are: 2SO $\rightarrow$
  (SO)$_2$, and SO $+$ (SO)$_2$ $\rightarrow$ S$_2$O $+$ SO$_2$
  (Schenk and Steudel, 1965). At low temperatures (i.e., in-eclipse),
  both SO$_2$ and S$_2$O condense out, and SO may effectively condense
  out through chemical reactions in the gas-phase with the surface,
  producing the above mentioned compounds (Hapke and Graham, 1989).
  Based on our observations, it looks like such self-reactions of SO
  must be very fast. Although this possbility has been suggested in
  the past (e.g., Lellouch, 1996), the SO ``condensation'' rate has
  never before been observed.}

\added{As shown,} the connection with volcanoes\deleted{, however,} is \deleted{much} more
tenuous for the SO emissions \added{than for SO$_2$}, except perhaps for
Daedalus Patera. \added{However, as noted above, SO's column density is not
really expected to change much. With a layer of SO, and perhaps
other noncondensible species (e.g., O, O$_2$), SO$_2$ may indeed not
completely collapse, such as modeled by e.g., Moore et al. (2009). We may also see emissions from stealth volcanoes, as
postulated by de Pater et al. (2020) to explain} 
\deleted{This reminds us of} the widespread spatial distribution of the 1.707 $\mu$m SO emissions,
which only occasionally showed a connection to volcanoes. \deleted{and was
postulated to originate from stealth plumes (de
Pater et al., 2020).} \deleted{It is We suggest that the SO, as well as some of the SO$_2$ not
obviously connected to volcanic sites, may also result from stealth
volcanism.}

\paragraph{Disk-integrated flux densities:}

We integrated the flux density over Io on each map for each
transition (except spw7, the lowest line strength, where we have no
detection), and plotted the results on the left panel of Figure~\ref{fig:ec_sun}. For
easier comparison, all flux densities in this figure have been
normalized to a geocentric distance of 5.044 AU, at which distance
Io's diameter is 1''.
Assuming that Io's flux density is constant
while in-sunlight, it decreases exponentially within the first
few minutes after the satellite enters eclipse. The dotted lines show
the collapse for each transition, modeled for the SO$_2$ lines as:

\begin{equation}
  F_i = F_i(t_{0}) e^{-t_j/(H_i+C_i(t_j-t_{1}))}
\end{equation}

  \noindent where $F_i$ stands for the flux density in each transition
  $i$ (for $i=$ spw1, spw2, spw5, and spw6), $t_j$ the time (in min)
  from time $t_{0}=0$ taken as midway during the partial eclipse\deleted{
  to $t_{3}$ 20 min later}\added{, $t_1=$8, $t_2=$ 10.7, $t_3=$ 19.5
    min).}  $H_i$ shows the exponential decay constant
  in minutes (indicated on the
  figure). After the initial drop
  in intensity, further decrease is slowed down, as roughly indicated
  by the term $C_i(t_j-t_{1})$, with $C=1$ at 346.524 and 332.091 GHz
  (spw1, spw5), $C=0.6$
  at 346.652 GHz (spw2), and $C=0.7$ at 332.505 GHz (spw6). A new
``steady state'' appears to be reached within $\sim$20\deleted{--40} minutes, similar to the results shown
by Tsang et al. (2016) at 19 $\mu$m. The flux density decreases by a
factor of 2 at 346.524, 332.091, and 332.505 GHz (spw1, spw5, spw6),
and by 3.2 at the strongest line transition, 346.652 GHz (spw2).
As shown by $H_i$, the latter flux density decreases much
faster than the others.
Also the SO emission, plotted here as the average of the two
transitions, decreases by a factor of 2, although much
more gradual, essentially following a linear 
decay, modeled as:

\begin{equation}
  F_i = F_i(t_{0}) + a t_j 
\end{equation}

\noindent where $a=-15$ mJy/min.
This more gradual decrease is also visible on the maps in
Figure~\ref{fig:MarchSO2}. 

\subsubsection{Line Center Maps on 2 \& 11 September 2018}\label{sec3a.2}

\paragraph{2 Sep. 2018}

  Figure \ref{fig:Sep2SO2} shows the distribution
of SO$_2$ and SO gases on 2 September 2018, when Io moved from eclipse
into sunlight. The integrated flux densities are plotted on the
right-side panel in Figure~\ref{fig:ec_sun}. As soon as sunlight hits
the satellite, SO$_2$ starts to sublime and within 10 minutes the
atmosphere has reformed in a linear fashion. The flux density in each
transition increased by roughly a factor of 2. SO
increased by roughly a factor of 3, also in a linear fashion. The
dotted lines on the figure were calculated using equ. 2; the values
for $a$ in mJy/min (positive sign for increasing slope) are indicated on the figure.

When the satellite was in-eclipse on 2 September (scan 6a in
Fig. \ref{fig:Sep2SO2} ), the spatial distribution of SO$_2$ gas shows
very strong emission near the SW limb, centered on P207 (91$^0$W
long., 37$^0$S lat.), a small dark-floored patera. Although thermal
emission has been detected at this site with the W. M. Keck
Observatory (Marchis et al., 2005; de Kleer and de Pater, 2016), no
evidence of plume activity has ever before been recorded. Faint emissions can
further been seen near Nyambe Patera and just north of PFd1691, a
dark-floored patera where thermal emission has also been detected with
the Keck Observatory (de Kleer et al., 2019).  As soon as Io enters
sunlight (scan 6b), the SO$_2$ emission near P207 becomes more
pronounced; this is the side of Io where the Sun first strikes. Over
the next 4--5 minutes (scans 8a, 8b) the emissions get stronger, in
particular near the volcanoes. At 9 minutes (scan 12) the SO$_2$
atmosphere has completely reformed.

The bottom row in Figure \ref{fig:Sep2SO2} shows practically no SO
emissions while Io is in eclipse (scan 6), except for some emission along
the limb north and south of P207. Faint
emissions are also seen near PFd1691, and at a few other places on the
disk. None of these emissions seem to be directly associated with
known volcanoes, nor with the SO$_2$ emissions on Io-in-eclipse.
About 4 minutes after entering sunlight (scan 8) strong SO
emission is detected above P207, suggestive of formation through photodissociation of SO$_2$.
Five minutes later we also detect
emissions over Nyambe Patera, and another 20--30 min later the
SO emissions track the SO$_2$ emissions pretty well, as expected if
the main source of SO is photolysis of SO$_2$.

\paragraph{11 Sep. 2018}

Figure \ref{fig:Sep11SO2} shows the spatial distributions of SO$_2$
and SO of Io-in-eclipse and in-sunlight on 11 September 2018, but not
during the transition from eclipse into sunlight. While in-eclipse, faint volcanically
sourced SO$_2$ emissions are present near P129, Karei and Ra Paterae,
and along the west limb near Gish Bar Patera and NW of P207. The
eruption at P207, so prominent 9 days earlier, has stopped. No SO
emissions are seen above the noise level. Ten minutes later, the
atmosphere has reformed as shown by the in-sunlight map, with most of
the emissions confined to latitudes within $\sim |30^0-40^0|$, in
agreement with the latitudinal extent measured from UV/HST data (Feaga et
al., 2009), and with Figures \ref{fig:MarchSO2} and \ref{fig:Sep2SO2}. The
SO map shows emission peaks above Karei Patera and P129.  The ratio of
flux densities between in-sunlight and in-eclipse is about a factor of
4--5 for SO$_2$ and $\sim$10 for SO on this day
(Fig.~\ref{fig:ec_sun}). Hence, as shown both by this large ratio and
the maps, on this date there was not much volcanic activity.

\subsubsection{Map of KCl on 20 March 2018}\label{sec3a.3}

On 20 March 2018 we also detected KCl, shown in Figure
\ref{fig:MarchKCL}. As shown, the distribution is completely different from that seen
in SO$_2$ and SO: the south-eastern spot is centered near Ulgen
Patera, \added{and emission is seen along the limb towards the north.} \deleted{there is a second source north-east of Pele, near Bo\"osaule
Montes, the highest mountain on Io, and a third source} \added{There
may also be some emission from} near Dazhbog
Patera.  \deleted{No difference was detected between the
in-sunlight and in-eclipse data sets.} KCl was not detected in September
2018, when \deleted{Pele and} Ulgen Patera was out of view. Further analysis of
the KCl data will be provided in a future paper.

\subsection{SO$_2$ Line Profiles and Image Data Cubes (Resolution
  0.142 km/s $\approx$ 160 kHz)}\label{sec3b}

In addition to the spatial distribution at the peak of the line
profiles (i.e., the line center maps) when Io goes from sunlight into eclipse and vice versa, the
full image data cubes contain an additional wealth of information.

\subsubsection{Image Data Cubes on 20 March 2018}\label{sec3b.1}

In Figure
\ref{fig:MarchFrames} we show several frames of the SO$_2$ image data
cubes together with disk-integrated line profiles from March 2018 for
Io-in-sunlight (top half) and in-eclipse (bottom half). To
increase the SNR, we averaged the data at 346.524 and 346.652 GHz (spw1 and spw2). We
also averaged all scans for the in-eclipse data (Table 2, scans 7--15
in Set 2) in this view.

At the peak of the line
profiles (frame 3), the images look similar to those shown in
Figure \ref{fig:MarchSO2}.  Moving away from the peak we see the
SO$_2$ distribution at a particular radial velocity, $v_r$ (the
velocity along the line of sight). It is striking how similar the
images are moving towards lower
or higher frequencies (positive or negative $v_r$), i.e., the brightness distribution is very symmetric
around the peak of the line. 
If there would 
be a horizontal wind of $\sim$300 m/s in the prograde direction, as
reported by Moullet et al. (2008) from maps 
when Io was near elongation (i.e., a different viewing geometry), we
would expect spatial distributions asymmetric with respect to the line center.
In the first frame, where we see the spatial distribution of gas
offset by $\sim +$0.6 MHz, i.e.,
moving towards us (blue-shifted -- B) at a speed of $\sim$0.5 km/s
($v_r=-0.5$ km/s), we would expect SO$_2$ gas on the west (left)
limb; we would see the gas on the east limb in
frame 5 where we map the brightness distribution at $v_r=+0.6$
km/s. On frame 2 emission would be concentrated on the west
hemisphere, and on frame 4 on the east hemisphere. The data show very
different spatial distributions. In
addition, on frame 5 we see faint emissions on both limbs, i.e,
material moving away from us on either side of the satellite, such as
expected for day-to-night flows. There is also faint blue-shifted
emission along both limbs in frame 1. 
Hence emissions due to a prograde wind cannot be distinguished in
these data.

On frames 1 and 5, both in-sunlight and in-eclipse, emission
from \added{near} Daedalus Patera dominates. This emission also dominates on frames
2--4 in-eclipse, and is clearly visible in-sunlight as well. Emission
from the vicinity of Karei Patera is
also visible on frames 2--4 in-eclipse and in-sunlight, as well as on
frame 1 in-eclipse. Emission may additionally originate
near N. Lerna in several frames. The line profiles,
in particular the high velocity wings, are clearly
dominated or produced by the volcanic plumes.

\subsubsection{Image Data Cubes on 2 \& 11 September 2018}\label{sec3b.2}

Figure \ref{fig:Sep2Frames} shows the image data-cube from September
2 when Io moves from eclipse into sunlight. The top half shows the
image data cube when Io was in sunlight, and the bottom half shows the
results for scan 6 when Io was in eclipse. As for the March data, the
broad asymmetric wings of the line profile are clearly produced by 
volcanic plumes, the plume at P207 on this date. 

On 11 September the situation is slightly different, as shown in Figure
\ref{fig:Sep11Frames}. There were no detectable volcanic plumes. When
Io was in-eclipse, only faint SO$_2$ emissions were seen. \added{At
  the peak of the line, emissions seem to originate near Euboea
  Fluctus and Ra Patera.} \replaced{Since}{But overall, if} SO$_2$
\replaced{must be}{is} volcanically sourced, most faint emissions may be sourced from stealth
volcanism, as mentioned in Section~\ref{sec3a}.
On the sunlit image data cube we
see some emission from the west limb in frame 1, near Zal Patera
(northern spot) and Itzamna Patera (southern spot), and on the east limb on
frame 5 at Mazda Patera. In frame
2 the emission has shifted more towards the center of the disk, but is
still only visible on the western hemisphere, i.e., the side that is moving towards us. In frame 4
more emission is coming from the eastern hemisphere, while in frame 5
emission comes primarily from the eastern limb.
These frames could be interpreted as indicative of a $\sim$300--400 m/s prograde zonal wind (i.e., on top
of the satellite's rotation around its axis), although it is
not clear why it would be offset from the equator in frame
1. Moreover, such a prograde zonal wind would result in line
profiles that are broader than those observed, even when modeled with
an atmospheric temperature of $\sim$~145 K.
Clearly, the
spatial distribution on this day is not as symmetric around the
center of the line (frame 3) as on the other two dates. On this date
most SO$_2$ must have been produced by sublimation, since we do not
see clear evidence of volcanic eruptions in-sunlight nor in-eclipse. 
This may be the
reason that we can distinguish zonal winds such as reported before by Moullet et
al. (2008). If these winds are real, they must form within 10--20 minutes
after Io re-emerges in sunlight. The reason for zonal winds, if indeed present, remains a
mystery, since we would expect day-to-night winds on a body with a warm
day- and cold nightside \added{(see, e.g., Ingersoll et al., 1985; Walker et
al., 2010; Gratiy et al., 2010)}.


\section{Analysis of SO$_2$ and SO Line Profiles}\label{sec4}

As shown by Lellouch et al. (1990), the SO$_2$ line profiles as
observed are saturated, and the peak flux density depends not only on the
temperature and column density, but also on the fraction of the
satellite covered by the gas. With our spatially-resolved maps and five
observed SO$_2$ transitions, we should be able to determine the
atmospheric temperature, column density and fractional coverage, as
well as constrain the presence of winds. This was
not possible with any of the previously published observations.

\subsection{Fractional Coverage.}\label{secfr}

The fractional coverage of the gas on Io can be determined directly
from maps of the SO$_2$ gas as observed in the various
transitions. However, the fractional coverage as seen on such maps
(Fig.~\ref{fig:MarchSO2}) is significantly affected by beam convolution, which
makes it hard to determine the precise fraction. A better way is to
use a deconvolved map \added{such as shown in Fig.~\ref{fig:cc} and
  discussed in Section \ref{sec2}}. \deleted{Such a map is stored as the
.model map in CASA's tCLEAN routine, which is composed of the sum of
all CLEAN components per pixel (using the Clark or H\"ogbom
algorithm,} The total number of pixels with
non-zero intensities divided by the total number of pixels on Io's
disk gives us the fractional coverage of the gas over the disk. \deleted{The maps shown in the
various figures are these .model files convolved with the gaussian
HPBW (Table 2), added to the residual map (i.e., the map left after
all clean components have been
subtracted; this map ideally is just
noise).} This procedure works best if the SNR in the maps is high, which is certainly
true for the strongest transitions, i.e., 346.652 GHz (spw2), and likely for SO in sunlight
(346.528 GHz, spw1), as shown in Figures \ref{fig:MarchSO2},
\ref{fig:Sep2SO2}, and \ref{fig:Sep11SO2}. The .model files cannot be trusted to accurately represent fractional SO gas coverage in-eclipse, since the signal is so low (hardly above the noise). If the brightness
distribution is very flat, like the continuum maps of Io, this
procedure underestimates the fractional coverage; it works best if the
spatial distribution consists of point-like sources.  
The fractional
coverage, $fr_{\rm map}$, for SO$_2$ as determined from maps in-eclipse and in-sunlight for
all 3 days is summarized in Column 3 of Table 3. We typically see a 30--35\%
coverage for Io-in-sunlight. On 20
March, $\sim$15 minutes after entering eclipse, we measured
$\sim$17\%, significantly higher than in September where we measured
$\sim$10\% when Io had been
in Jupiter's shadow for $\sim$2 hours. 
The fractional coverage in-eclipse \replaced{should}{may} primarily depend on Io's volcanic activity,
which may vary considerably over time. Interestingly, although there
was not much volcanic activity on 11 September, the fractional
coverage was quite similar to that measured on 2 September, when P207
was extremely active. This suggests a very vigorous eruption at P207,
but small in extent, essentially a point source in our maps.
For SO in sunlight we measured $fr_{\rm
  map} \approx $16\% in March and $\sim$10\% in September. In-eclipse 
this coverage drops to below $\sim$5\%, and cannot be measured
very acurately. We estimate an uncertainty of $\sim$10\% on all
retrieved numbers for SO$_2$, and $\sim$20\% for SO.

\subsection{Radiative Transfer Model}\label{secRT}

To model the line profiles, we developed a radiative transfer (RT) code analogous to that used
to model CO radio observations of the giant planets
(Luszcz-Cook and de Pater, 2013). We assume Io's atmosphere to be in
hydrostatic equilibrium, so the density can be calculated as a
function of altitude once a temperature is chosen (we use an isothermal
atmosphere in this work). Any molecular emissions are assumed to occur
in local thermodynamic equilibrium (LTE), as expected for these
rotational transitions in Io's atmosphere (Lellouch et al., 1992). We perform RT calculations across Io's disk at a cell size of 0.01''
and a frequency resolution 1/4th of the resolution in the
observations (i.e., roughly 40 kHz). Io's solid body rotation
($v_{rot} =$ 75 m/s at the equator) is
taken into account; a simple increase/decrease in $v_{rot}$ can
account for zonal winds.

In order to account for potential Doppler shifts (blue- and
red-shifts) in line profiles, which might be expected for localized volcanic
eruptions or for day-to-night winds in disk-averaged line profiles, we added a separate parameter,
$v_r$, in addition to the planet's rotation and zonal winds. With this
parameter we can accurately fit any offset in frequency at line
center. As shown below, we do need the freedom to shift some modeled line
profiles to match the data; potential reasons for such shifts are
discussed below and in Section 5.

We
adopted a surface temperature of 110 K with an emissivity of 0.8. For the analysis of our data we ran many models,
where we varied the column density, $N$, from $\sim 10^{15}$ -- few
$\times 10^{17}$ cm$^{-2}$ for SO$_2$, a factor of 10 smaller for SO,
the temperature $T$ from$\sim$ 120 -- 700 K, and the rotational and
Doppler velocities, $v_{rot}$ and $v_r$, each from $\sim -400$ to $+$400 m/s.  In the
following subsections we analyze line profiles for March and
September.

\added{Figure~\ref{fig:contr} shows sample contribution functions for the
four SO$_2$ line transitions detected in our data. The line profiles
based upon the parameters in
panel a) match the observed line profiles quite well, as shown in Sections
\ref{sec4a} and \ref{sec4b}. Panels c and d show the changes in the
contribution functions when the temperature or column density are
changed. Line profiles based upon these parameters do {\it not} match
our observed line profiles, but give an idea where one probes under
different scenarios. The column density used in panel d) matches that
usually reported for the anti-jovian hemisphere. while the temperature
in panel c) is similar to the atmnospheric temperature determined at 4
$\mu$m (Lellouch et al., 2015). In panel b we show a
calculation for a temperature that increases with altitude, such as
expected for Io based upon plasma heating from above (e.g., Strobel et
al., 1994; Walker et al., 2010). Resulting line
profiles again do {\it not} match any of our data. The bottom line of
this excercise is that we typically probe the lower 10 up to $\sim$80
km altitudes for column densities of $\sim$10$^{16}$--10$^{17}$
cm$^{-2}$, and that different transitions are sensitive to
different altitudes in the atmosphere. We further note that the
temperature structure in the first few tens of kilometers above the surface is unknown,
which makes interpretation of mm-data quite challenging.}

\subsection{SO$_2$ on 20 March 2018: Sunlight $\rightarrow$ Eclipse}\label{sec4a}

\vskip 0.15truein

\subsubsection{Disk-integrated line profiles.}\label{sec4a.1}

We first focus on the disk-integrated line profiles of SO$_2$ for Io-in-sunlight. We have 5
transitions; although there essentially is no signal in the weakest
line transition (333.043 GHz, in 
spw7), it still helps to constrain the parameters. The free parameters
in our model are $N_t$, $T_t$, $fr_t$, $v_{rot}$ and $v_r$, where the
subscript $t$ is used for disk-integrated data. We thus
have to find a set of parameters that can match the line profiles in
all SO$_2$ transitions. Moreover, since the fractional coverage, $fr_t$,
should match that derived from the maps, $fr_{\rm map}$ (Table
3), the parameter $fr_t$ is heavily constrained for disk-integrated
line profiles.

While the Doppler shift, $v_r$, in our implementation leads to a shift
in frequency (i.e., velocity) of the entire line profile, both the
temperature and rotation of the body (or any zonal wind), $v_{rot}$,
lead to a broadening of the line profile. Hence high values of
$v_{rot}$ can be compensated by lower atmospheric temperatures. For
example, for $v_{rot}=$300 m/s and $T_t=$195 K, the lineshape matches
the observed profiles quite well; however, for any given $N_t$, there
is not a single value for $fr_t$ that can match the line profiles for
all transitions; moreover, $fr_t$ should be equal to $fr_{\rm map}$.
Based upon such comparisons we can rule out zonal winds much larger
than $\sim$100 m/s, which agrees with our earlier findings where we
did not see evidence on the maps for large zonal winds\added{, except
  perhaps for September 11}
(Section~\ref{sec3b}).  Since there is no noticeable broadening in the
line profiles for zonal winds up to $\sim$100 m/s, we ignore any
potential presence of zonal winds in the rest of this section, and
simply use $v_{rot}=$ 75 m/s.

By assuming that the fractional coverage of SO$_2$ on Io, $fr_t$,
should be the same for all transitions, and be equal to
$fr_{\rm map}$, we get a pretty tight constraint on the column density
and atmospheric temperature: $N_t = (1.35 \pm 0.15) \times 10^{16}$
cm$^{-2}$ and $T_t = 270 \pm ^{50}_{25}$ K.
These numbers are summarized in
Table 3, together with $fr_t$; the spread in $fr_t$ between
transitions is written as an uncertainty.

We found
that the modeled profile had to be shifted in its entirety by $+$20 m/s
(22--23 kHz), with an estimated error of 7 m/s. Since uncertainties in the line positions as
measured in the laboratory are of
order 4 kHz\footnote{https://spec.jpl.nasa.gov}, the observed offset
cannot be caused by measurement errors in the lab. This shift is 
indicative of material moving away from us. This can be caused by an
asymmetric distribution of the gas with more material on the eastern
than western hemisphere. Alternatively, it can be caused by
day-to-night flows, or gas
falling down onto the surface such as expected in volcanic eruptions
after ejection into the atmosphere. The rising branch of gas plumes
usually occurs over a small surface area (vent), \added{is very dense
($\sim$few 10$^{18}$ cm$^{-2}$; see, e.g., Zhang et al., 2003), and
therefore saturated.}  While rising, the
plume cools and expands, and the return umbrella-like flow,
essentially along ballistic trajectories, covers a
much larger area, up to 100's km from the vent, \added{with column densities
about 2 orders of magnitude lower than at the vent}. Hence, since the downward
flow covers a much larger area than the rising column of gas, one can
qualitatively explain a redshift of disk-integrated line
profiles. \added{This idea was used by Lellouch et
  al. (1994; see Lellouch, 1996 for updates) to explain $\sim$80 m/s redshifts in their line
  profiles, which they could model if there would be of order 50
  plumes on the observed hemisphere. Although this seemed quite a
  large number of plumes at the
  time, if one considers the presence of stealt plumes (Johnson et
  al., 1995) and the recent publication of the spatial distribution of
  1.707 $\mu$m SO emissions (de Pater et al., 2020), this may be a quite
  plausible idea.}

Several fits are shown in Figure
\ref{fig:lineprofilesM}, panels a and c\deleted{(additional figures are provided in the
Appendix)}. For each of the models shown we used the
mean fractional coverage as
derived from the line profiles in the four spectral windows (spw1, spw2, spw5, and
spw6) for that particular model, i.e., $fr_t =$ 0.32 for the best-fit
model ($N_t = 1.35 \times 10^{16}$), but $fr_t =$0.39 for $N_t = 1 \times 10^{16}$
cm$^{-2}$ and $fr_t =$0.26 for $N_t = 2 \times 10^{16}$
cm$^{-2}$. While all three model curves might match one or two spectral windows,
only one curve (red one) fits all transitions, as well as $fr_{\rm map}$. Note, though, that none
of the curves fits the broad shoulders of the observed profiles; this
is clearly caused by the relatively high velocities (Doppler shift) of
the eruptions, as discussed in Section \ref{sec3b} and Figure
\ref{fig:MarchFrames}.

The column density and temperature hardly change for Io-in-eclipse (scan 11$+$15) (Table 3). The drop in
flux density is mainly caused by the factor of $\sim$2 drop in fractional
coverage. In other words, a smaller fraction of the satellite is
covered by gas, but over those areas the column density and
temperature are essentially the same as those seen on Io-in-sunlight. Line profiles are shown in panels b and d of Figure
\ref{fig:lineprofilesM}. We note the discrepancy between the data and
models  in panel d, indicative of shortcomings in our model: while the
line profiles of Io-in-sunlight can be modeled relatively well with
our simple hydrostatic model,
the model falls short when the gas emissions are dominated by
volcanic plumes rather than by SO$_2$ sublimation. In this particular case there appears to be 
excess emission at lower frequencies, i.e., at velocities moving
away from us.

\subsubsection{Line profiles for individual volcanoes.} 

We next investigate the line profiles of individual volcanoes, Karei
and Daedalus Paterae. These line profiles were created by integrating
over a circle with a diameter equal to the HPBW (Table 2) 
centered on the peak emission of the volcano on the 346.652 GHz (spw2)
map. We
determined the line profile for the models in the exact same way, so
that the rotation of the satellite was taken into account, and the
viewing geometry (i.e., pathlength through the atmosphere) was the same. Hence, the
modeled line profile for a volcano on the West (East) limb is already Doppler shifted
to account for the satellite's rotation towards (away from) us, and any additional shifts are
intrinsic to the volcano itself. 
As shown, the hydrostatic line profiles match the observed
spectra for Karei Patera in sunlight very well
(Fig.~\ref{fig:lineprofilesM}e) with a column density and temperature
that are
quite similar to the numbers we found for the integrated flux
densities, but with a fractional coverage of
almost 50\% (Table 4). Hence, the column density
(cm$^{-2}$) of SO$_2$ gas in-sunlight appears to be quite constant across Io over areas
where there is gas, i.e., over approximately 30--35\% of Io's surface
in-sunlight, and over about half the area of a volcanically active
source (note that we integrated here over approximately the size of the
beam, so the plume itself may be unresolved).

Since we cannot determine the fractional
coverage for individual volcanoes from the maps, we have to solely
rely on finding models that give us the same fractional coverage in
all four transitions. The uncertainty in $fr_v$ (the subscript $v$
stands for volcano) as listed in Table 4
shows the spread in $fr_v$ between the four transitions. If the spread
is small, the solution is quite robust. When the spread is
large, the results should be taken with a grain of salt. 
The line center is offset by $+$60 m/s,
indicative of material moving away from us, such as might be expected
for an umbrella-shaped plume as discussed above. The in-eclipse profile
(panel f) can
also be matched quite well, with a similar temperature, perhaps a
higher column density, but a much lower $fr_v$.

The observed profile for Daedalus Patera in-sunlight is very
different. The profiles in all four transitions have a pronounced red
wing (Fig.~\ref{fig:lineprofilesM}g). The main profile can be matched
quite well with $T_v \approx$ 220 K, and $N_v \approx$ 1.5$\times 10^{16}$
cm$^{-2}$, with a fractional coverage of 46\%. The line center appears
to be Doppler shifted by $-$40 m/s, i.e. material moving towards
us. Note that the line offsets for the two volcanoes are in the
direction of a retrograde, \added{rather than prograde, zonal} wind; however, if such a wind would prevail,
we would expect the windspeed to be largest near the limb (Daedalus Patera), i.e.,
opposite to the observations. \added{The observed Dopplershifts are
  more likely local effects, produced by the eruptions.}
For the in-eclipse profile (Fig.~\ref{fig:lineprofilesM}h) no good solution could be found, as indicated by
the large uncertainties.
This is not too surprising, since in-eclipse most emissions are likely volcanic in
origin, since as soon as SO$_2$ gas is cooled to below its condensation
temperature it \replaced{condenses out}{may condense out}. The applicability of our simple hydrostatic model
is therefore limited. \added{To properly model these one needs to add
  volcanic plumes to the model, such as done by e.g., Gratiy et
  al. (2010). (see also Section~\ref{sec5.3}).}
  \deleted{As shown, there is also clear evidence of high velocity
contributions, both positive and negative, which are not captured in
our simple model, though one could construct a model consisting of
different components to capture such high velocity wings. This,
however, is beyond the scope of this paper, but will be addressed in a
future publication (see also Section 5 on a comparison of our data
with volcanic plume simulations published by Zhang et al., 2003).}

\subsection{SO$_2$ on 2 \& 11 September 2018: Eclipse $\rightarrow$ Sunlight}\label{sec4b}

Figure~\ref{fig:lineprofilesS} shows several line profiles for the
September data\deleted{(additional line profiles are provided in the
Appendix)}; best fits are summarized in Tables 3 and 4.
As with the March data, the disk-integrated line profiles for both 2 and 11 September
for Io-in-sunlight can be modeled quite well with our simple hydrostatic
model, in contrast to line profiles taken of Io-in-eclipse where
emissions must be volcanic in origin, and the applicability of our
model is limited. From our hydrostatic models we find that the SO$_2$
fractional coverage on both days is a
factor of 3 lower for the in-eclipse data than for Io-in-sunlight, while it was only a factor
of 2 in March. On the latter date the satellite had only been in
shadow, though, for 15 minutes, much shorter than for the September
data. While on 11 September the column density between in-sunlight and
in-eclipse data is very similar, on 2 September it may be a factor of
2 higher when in-eclipse, although the uncertainties are large enough
to accommodate no-change as well.

Line profiles of individual volcanoes, calculated by integrating over
a circle with a diameter equal to the HPBW, also deviate significantly
from the hydrostatic models, although for volcanoes in-sunlight the
discrepancies are smaller than when in-eclipse. We 
were able to find a good model for P207, in particular in-sunlight,
where a column density quite similar to that found for the
disk-integrated line profiles covers $\sim$60\% of the volcano. During
eclipse the fractional coverage decreases by a factor of $\sim$2 (or
more), while the column density may not vary much (considering the
uncertainties). In contrast, even though the
observed line profiles for Nyambe Patera look quite gaussian both
in-sunlight and in-eclipse, we were
not able to find a model for either data set with the same $fr_v$ for all 4
transitions, which translates into a high uncertainty even for the
in-sunlight data.

\subsection{SO Line Profiles}\label{sec4c}

We modeled the disk-integrated SO line profiles in Figure \ref{fig:lineprofilesSO} by adopting the temperature that was
determined from the SO$_2$ profiles on the various days, since the atmospheric temperature
should not depend on the species considered. The fractional coverage
as determined from the line center maps for Io-in-sunlight is about a factor of 2
lower for SO than for SO$_2$ in March, and more like a factor of 3--4 in
September. 
The temperature together
with this fractional coverage should
result in a trustworthy value for the column density, assuming again
that the atmosphere is in hydrostatic equilibrium. With these
assumptions we find a column density of $\sim 10^{15}$ cm$^{-2}$ on 20
March when in-sunlight, roughly a factor of 10 below the SO$_2$
column density. This, with the lower fractional coverage, suggests a
difference of a factor of $\sim$20 between the total volumes of SO and
SO$_2$ gas\added{, in good agreement with previous observations (e.g.,
Lellouch et al., 2007)}.
On 2 September the column density is roughly a
factor of 5 lower than the SO$_2$ column density, which with the
much lower fractional coverage also suggests almost a factor of 20
difference in gas volumes. On 11 September the column density is 
again a factor of 10 below the SO$_2$ value, but with a much lower
fractional coverage this results in a difference of $\sim$40 between the
volumes of SO$_2$ and SO gases.

As shown in Figures \ref{fig:MarchSO2},
\ref{fig:Sep2SO2}, \ref{fig:Sep11SO2} and \ref{fig:lineprofilesSO} we
did detect SO when Io was in-eclipse in March and on 2 September, but
not on 11 September. The SNR in the maps, however, is very low, which
prevented a good estimate of the 
fractional coverage, a necessary quantity to determine the column
density from the data. Assuming the same temperature as derived from
the SO$_2$ maps for Io-in-eclipse, we find a fractional coverage
of 7\% for an SO column density of 10$^{15}$ cm$^{-2}$, i.e., about
half the fractional area for the same column density as seen in the
in-sunlight maps. The fractional coverage decreases for a higher
value of $N_t$, and vice versa for a lower value. On 2 September the
most likely scenario for Io-in-eclipse is that both $fr_t$ and $N_t$ decrease by a factor
of 2, while on 11 September no SO emissions were detected in-eclipse.

\section{Discussion}\label{sec5}

\subsection{Summary of Observations}\label{sec5a}

We observed Io with ALMA in Band 7 (880 $\mu$m) in five SO$_2$ and two SO
transitions when it went from sunlight into eclipse (20 March 2018),
and from eclipse into sunlight (2 and 11 September 2018). On all three
days we obtained disk-resolved data cubes, and analyzed SO$_2$ and SO
line profiles for both the disk-integrated data, and for several active volcanoes on
the disk-resolved data cubes. Specifics on the observations and derived
parameters are summarized in Tables 1--4.

\subsubsection{Disk-integrated data: SO$_2$}\label{sec5a.1}

The line-emission disk-integrated flux
densities\footnote{The disk-integrated flux densities are normalized
  to a geocentric distance of 5.044 AU for intercomparison of the datasets.} are typically $\sim 2-3 \times$ higher for Io-in-sunlight
than in-eclipse (Fig.~\ref{fig:ec_sun}), indicative of a roughly 30--50\% contribution
of volcanic gases to the SO$_2$ emissions. However,
there is much variability in these numbers. On 11 September the SO$_2$ flux
density in-sunlight is 4--5 times higher than in-eclipse.  In March, when Io went
into eclipse, the flux density in the strongest transition, $F_{\rm 346.652}$, dropped exponentially by a factor of 3, in
contrast to the factor-of-2 drop in the three weaker transitions. $F_{\rm 346.652}$
for Io-in-sunlight was $\sim 2 \times$ higher than 
$F_{\rm 346.524}$, about 30\% above the ratio in their intrinsic line strengths (Table 1). In
contrast, the observed ratios between $F_{\rm 346.524}$ with $F_{\rm
  332.091}$ and with $F_{\rm 332.505}$ are 40 and 60\% smaller than the ratios between
their intrinsic line strengths. Since the flux density in the various
transitions depends also on the atmospheric
temperature, which determines (in LTE) which energy levels in the
molecule are
populated (Boltzmann's equation), the differences in flux density
between the various transitions was used in Section~\ref{sec4} to
determine the column density and atmospheric temperature. 
For example,
for lower temperatures, the modeled $F_{\rm 346.652}$ would be too low, while
the modeled $F_{\rm 332.505}$ would be too high\added{, which can be
  qualitatively understood from the difference in contribution
  functions between panels a and c in Figure~\ref{fig:contr}}. We found that neither the atmospheric
temperature nor the column density between the in-sunlight and in-eclipse data sets did 
noticeably change, but that the differences in flux density could be accounted
for by a factor of 2--3 decrease in fractional coverage. Or in other
words, the column densities (cm$^{-2}$) remained the same, but there
were fewer areas (2--3 times less) above which SO$_2$ gas was present.

Tsang et al (2016) measured a factor of 5$\pm$2 drop in column density when Io moved
from sunlight into eclipse. However, since they cannot distinguish
between a high column density with low fractional coverage and a low
column density with a high fractional coverage, our findings
essentially agree.

All observations could be matched quite well with 
an isothermal atmospheric
temperature of 270$\pm$50 K. This is clearly an oversimplification,
since the temperature 
\added{will certainly vary with altitude, and also with latitude, longitude, and time of day. Walker et
  al. (2010) show that the atmospheric (translational) temperature
  rises steeply with altitude due to plasma heating from above. Near
  the surface the SO$_2$ gas is expected to be in equilibrium with the
  surface frost, rising to $\sim$400 K at an altitude of 70 km during
  the day; at night the plasma can reach lower altitudes so that the
  400 K temperature may be reached at an altitude of
  $\sim$40 km. The exact 3D temperature profile depends on the
  3D distribution of the atmospheric density, which for SO$_2$ is
  tightly coupled with the frost distribution and temperature.}
\deleted{ One might expect the temperature to decrease with altitude
for Io-in-sunlight, as in adiabatic profiles in a planet's
troposphere.} \added{Moreover, as seen from the previous sections,
volcanic plumes may affect the atmosphere and its temperature
structure quite dramatically. We will get back to this in Section~\ref{sec5.3}. The bottom
line is that altitude-dependent
changes in temperature and density affect the various transitions
in different ways, as shown by the contribution functions in Figure~\ref{fig:contr}.} \deleted{Moving into eclipse,
the upper levels cool down much faster, and perhaps more, than levels near the ground,
which could qualitatively explain the larger drop in $F_{\rm 346.652}$
compared to that at the other frequencies. Experiments with non-isothermal
atmospheres, however, is beyond the scope of this paper.} 

In contrast to the exponential decrease in intensity during eclipse
ingress, there is a linear increase during egress for at least about 10
minutes. Interestingly, the
SO$_2$ flux density in three of the four transitions $\sim$ 10 minutes after emerging from
eclipse on 2 September is higher than the values 1/2 hr later:
4$\pm$2.6\% higher for $F_{\rm 346.652}$, 16$\pm$7\% for $F_{\rm 332.091}$
and 19$\pm$7\% for $F_{\rm 332.505}$.

The disk-integrated flux densities
for September 11 are shown alongside the September 2 numbers in Figure
\ref{fig:ec_sun}. The in-eclipse flux densities for SO$_2$ on Sep. 11 are typically a factor of
2 (two strongest transitions) -- 3 (two weakest transitions) below the Sep. 2 in-eclipse
values. As shown in the line center maps, on Sep. 11 the volcanic
acitivity was very low\added{, which may explain the difference in
  flux densities between these dates}. When in sunlight, the Sep. 11 numbers for
SO$_2$ are \added{well below the Sep. 2 values,}  \deleted{2--12\% below the Sep. 2 values $\sim$40 min
after entering sunlight. It may be more appropriate to compare the Sep. 11 data with the
Sep. 2 data $\sim$10 min after entering sunlight (Table 2), when the
difference is even higher (9--30\%);} in
particular for the
two weakest transitions, which are lower by a factor of 1.18$\pm$0.09
at 332.091 GHz, and 1.29$\pm$0.09
at 333.043 GHz, compared to the nearby high Sep. 2 values.

\deleted{Since all SO$_2$ at temperatures below its dew point condenses out, only gas
volcanic vents and warm surface areas remains is pesent on Io-in-eclipse. These
warm volcanic areas apparently cover $\sim$10\%, and at times up to almost 20\% of
Io's surface. In some cases there is a clear connection to a
particular eruption; in other cases we suggest the presence of stealth
volcanoes, which presence has been postulated before on the
sub-jovian hemisphere (de Pater et al., 2020); also point sources
and glows of gases have been seen on this hemisphere with the New Horizons spacecraft, though they were interpreted in a
different way (Spencer et al., 2007).} 

\added{Since all SO$_2$
at temperatures below its dew point (i.e., near the surface) condenses out, only gas
sourced from
volcanic vents, or SO$_2$ gas that was prevented from complete collapse by a
layer of non-condensible gases (e.g., Moore et al., 2009) can be
present on Io-in-eclipse.
These gases apparently cover $\sim$10\%, and at times up to almost 20\% of
Io's surface. In some cases there is a clear connection to a
particular volcanic eruption; in other cases emissions from
volcanically-sourced gases could be caused by stealth
volcanism, which presence had been postulated to explain the 1.707 $\mu$m forbidded emissions on this
hemisphere (de Pater et al., 2020). Point sources
and glows of gases that were seen on this hemisphere with the New
Horizons spacecraft, interpreted as being caused by plasma interactions
with the (near-)surface (Spencer et al., 2007), could also be a
signature of stealth volcanism (de Pater et al., 2020).}

\subsubsection{Disk-integrated data: SO}\label{sec5a.2}

The SO flux density for Io-in-sunlight is highest on Sep. 2
(0.66$\pm$0.03 Jy) and lowest on Sep. 11 (0.45$\pm$0.04 Jy).
\added{Despite the fact the SO is not expected to significantly
  decrease during an eclipse (Section~\ref{sec3a}),} \deleted{The SO
flux density decreases when Io enters eclipse, but not as rapidly
as SO$_2$;} we see a gradual (linear) decrease by a factor of 2 in March.
Since SO does not
condense at these temperatures, \added{it is likely} \deleted{but may be} removed from the atmosphere
through reactions with itself on the surface, \added{at a much faster
  rate than hitherto anticipated (Section~\ref{sec3a}). \deleted{(Lellouch, 1996), one
might indeed expect a more gradual, rather than exponential, decrease.}
SO similarly is restored much more slowly than SO$_2$,
indicative of formation from SO$_2$ through photolysis. According to
Moses et al. (2002), SO is formed
through photolysis of SO$_2$ at a column abundance rate of $4.6\times
10^{11}$ cm$^{-2}$ s$^{-1}$; i.e., it takes about 1/2 hr to produce a
full column of $\sim$10$^{15}$ cm$^{-2}$ s$^{-1}$, \added{and less depending
on how much SO gas is left. Above
  volcanoes,} \deleted{They show
that} about 50\% of SO is produced this way, and}
another 50\% \added{may be produced at a similar rate} through the reaction of O$+$S$_2$. The data (Figure
\ref{fig:ec_sun}) show that \added{SO is fully restored} \deleted{a full column is created} within $\sim$10
min of time, which given the uncertainties in the ALMA column densities, and the
various processes to produce SO (Section~\ref{sec3a}), agrees pretty
well with the models. \deleted{ which given that only half a column needs to be created
from SO$_2$ photolysis agrees more or less with the models, in
particular since usually not all the SO was removed during eclipse.}

\subsubsection{Disk-resolved data and line profiles}\label{sec5a.3}

On 20 March the SO$_2$ emission is dominated by the volcanically
active Karei and Daedalus Paterae, while some low-level emission is
seen near North Lerna. On 2 September the emission is dominated by 
P207, while we also see emission near PFd1691 and Nyambe Patera. The
situation is less clear on 11 September: P207 Patera was no longer
active, while emissions on Io-in-sunlight were seen above Karei and
Nyambe Paterae. In-eclipse very low-level activity was seen over Karei,
Gish Bar, Ra and P129 Paterae, north of P207, and near Euboea Fluctus. SO emissions in
March tracked the SO$_2$ emissions reasonably well, i.e., both Karei
and Daedalus Paterae showed activity. On 2 September no clear SO
emissions were seen during eclipse, but $\sim$4 minutes after Io
emerged from eclipse SO emissions were detected over P207 and near Nyambe
Paterae. On 11 September no SO emissions were detected in-eclipse;
when Io was in-sunlight SO emissions were highest over Karei Patera,
and visible above P129, and faintly across the equatorial band,
more or less coinciding with the SO$_2$ emissions. This all suggests
that the main source of SO is photolysis of SO$_2$. \added{As mentioned
above, at volcanic eruption sites a full column of $\sim$10$^{15}$
cm$^{-2}$ s$^{-1}$ will be produced in 15 min. The SO
peak intensity levels above P207 changed from $\lesssim$0.04
Jy/bm in-eclipse (sc.6, Fig.~\ref{fig:Sep2SO2}), to $\sim$0.07 Jy/bm (sc. 8)
to $\sim$0.12 Jy/bm in-sunlight. Hence in $\sim$4 min
photochemical reactions likely produce enough SO to explain the
observations.} \deleted{and the onset of
photolysis is fast, on timescales of minutes.}

The effect of volcanoes on the SO$_2$ and SO emissions is most clearly
seen in the lineshapes. While disk-integrated Io-in-sunlight data can
usually be matched quite well with hydrostatic models, the in-eclipse
profiles deviate considerably from such gaussian-shaped profiles. Both
in-sunlight and in-eclipse disk-integrated line profiles show broad
low-level wings out to $\sim$1 MHz from line center, indicative of
velocities of order 800--900 m/s both towards and away from the observer. As shown on Figures
\ref{fig:MarchFrames} and \ref{fig:Sep2Frames}, these wings are
clearly caused by volcanic plumes.

The line profiles for individual volcanoes often show red- or
blue-shifted wings, while on some occasions the entire profile seems
to be shifted towards lower or higher frequencies. In particular in
March, the SO$_2$ line profile of Karei Patera showed a $+$60 m/s shift, and
a $-$40 m/s shift for Daedalus Patera, while the disk-integrated line profiles were redshifted
by $+$20 m/s. Based upon our earlier discussions, the redshifts may be
caused by the plume above Karei Patera, where ejection of
the gas is confined to a small (unresolved) area, while the plume
material falls back down on the surface over a much larger area, 100's of
km in radius, resulting in a redshift in the line profile. Daedalus Patera,
showing a blueshift (which in part offsets the redshift in the
disk-integrated line profile), must be dominated by material moving
towards us. Daedalus, in contrast to Karei Patera, is located very
close to Io's limb (Fig.~\ref{fig:MarchSO2}), and hence is seen under a
very different viewing geometry such that both the rising plume
material, but also part of the umbrella-shaped plume, is moving
towards us, resulting in a blueshift of the entire profile. The
redshifted wing on the volcano's line profile is still indicative of
material falling down onto the surface, away from us, much of it perhaps over the
limb.

The emission over P207 is also on the limb of the satellite, but here
we see a wing of blueshifted material while there is no noticeable
offset of the entire profile. Clearly, both the viewing geometry and
the exact geometry or shape of the plume and ejection itself\added{,
  in combination with overall windflow patterns (e.g., Gratiy et al., 2010),} are
important parameters that affect the line profile.

\subsection{Post-eclipse Brightening}\label{sec5b}

As mentioned above, the flux density on 2 September $\sim$10 minutes
after emerging from eclipse was considerably higher, up to $\sim$20\%
in some transitions, than 1/2 hr later. This appears to be an
anomalous post-eclipse brightening effect.  A $\sim$10\% brightening of the
satellite for about 10--20 min after emerging from eclipse was first
reported by Binder and Cruikshank (1964) at a wavelength of 450
nm\added{, i.e., they observed the satellite's surface in reflected
  sunlight}.
The authors noted that Io was $\sim$10\% brighter when it emerged
from eclipse, which decreased over the next $\sim$15 minutes. They
suggested that the brightening might be caused by an
atmospheric component that condenses on Io's surface during the
eclipse. This makes the satellite bright; the ice should evaporate
only minutes after receiving sunlight again, resulting in
a slow darkening, back to its original reflected-sunlight
intensity. We note that these observations were obtained before Io's
atmosphere and its volcanic activity were detected -- in fact, based on their
data, the authors suggested Io to have an atmosphere. During subsequent
years, both detections and non-detections (e.g., Cruikshank et al., 2010;
Tsang et al., 2015) of this ``post-eclipse
brightening'' effect have been reported at wavelengths from the UV to the
mid-IR. Explanations of the effect range from condensation with
subsequent sublimation of SO$_2$-frost (Binder and Cruikshank, 1964;
Fanale et al., 1981; Belluci et al., 2004), to changes in Io's reflectivity due to sulfur
allotropes as a result of changes in surface temperature (Hammel et
al., 1985), to interactions of amospheric molecules with Jupiter's
magnetospheric plasma (Saur and Strobel, 2004). Some authors
concentrate on phenomenae causing a brightening of the surface, and
others of the atmosphere. No clear explanation has been provided yet;
and as shown by the data, the effect has not always been detected,
which has been interpreted by possible differences in frost coverage at
different longitudes.

The gradual increase in flux density in our data during the first $\sim$10
minutes after emerging from eclipse into sunlight is exactly how
Binder and Cruikshank (1964) explained their observed post-eclipse brightening of
Io's surface: The surface was bright since the
SO$_2$-ice coverage had increased due to condensation while in
eclipse; as soon as the surface warmed, SO$_2$ sublimed, the
satellite's surface darkened, and the atmosphere reformed. The
situation, as we observed it, is a bit more complex in that we see
the
SO$_2$ flux density to ``overshoot'' after 10 minutes (the end of our
observing session 1 on Sep. 2), before reaching a steady state (in observing set
2 on Sep. 2). In the next section we show that this may result from the
interaction of volcanic plumes with the reforming atmosphere.

\subsection{Comparison of Data with Atmospheric Models}\label{sec5.3}

\subsubsection{Summary of Published Plume Simulations}\label{sec5c}

In the following we compare the above results with simulations of
volcanic plumes by Zhang et al. (2003) \added{and McDoniel et al. (2017)}. These simulations include
a full treatment of gas dynamics, radiation (heating and cooling
through rotational and vibrational radiation), sublimation and
condensation. \added{McDoniel et al. (2017) coupled Zhang et al.'s
  (2003, 2004) original
plume model to a model of a sublimation-driven atmosphere, developed
over the years 
by Moore et al. (2009) and Walker et al. (2010, 2012). 
Simulations with this coupled model show} \deleted{discussed above, to show} how a volcanic plume
on the dayside expands in a sublimating atmosphere. The authors present models for a Pele-type plume both on the
night and day side. They assumed a night side surface temperature of
90 K, and 116-118 K during the day. The gas erupts from the vent at a
temperature of $\sim$600 K and a source rate of $\sim$10$^{29}$ SO$_2$
molecules/second at hypersonic velocities of close to 1 km/s. It then expands and
cools. At an altitude of $\sim$300 km a canopy-shaped shock forms
\added{(due to Io's gravity field) }
where the radially expanding molecules turn back down to the surface.
\deleted{(analogous to ballistic trajectories).} Most of the gas falls down
$\sim$400--600 km from the vent. \added{On the night side, the SO$_2$ gas} \deleted{where it} condenses and forms a ring
around the volcano which matches the red ring observed around Pele. Due to
plume expansion and vibrational cooling the gas temperature above the
vent decreases to very low ($\sim$50 K) temperatures, while the
temperature in the canopy-shock is of order 300--400 K.

\deleted{At a distance
of 350 km and altitude of 100 km, the gas is nearly collisionless,
non-LTE, and the molecules show a bimodal distribution, with one set
mostly falling down to the surface at $v_r = 700$ m/s, and another
one moving radially outwards at 1100 m/s}

On the night side the model shows an average column density $N_v= 1.1
\times 10^{16}$ cm$^{-2}$ over a region up to $\sim$600 km from the
vent. Directly above the vent, though, $N_v \approx 10^{18}$ cm$^{-2}$, and
drops by an order of magnitude over a 30-km distance.

A Pele-type plume on the day-side is different because there is also
SO$_2$ sublimation from Io's surface, and hence the plume expands in a
background atmosphere.  \added{The extent to which a dayside
  sublimation atmosphere is affected by plumes depends on the size,
  density and ejection velocity of the plume, as well as on the
  density of the sublimation atmosphere, which is set by the
  temperature of the surface frost (for details, see McDoniel et
  al. 2017). Plumes that do not rise up above the exobase (i.e., the
  altitude at which the mean free pathlength between collisions is
  equal to one atmospheric scaleheight, which is typically at an
  altitude of $\sim$30--50 km on Io -- McDoniel et al., 2017), will
  not affect the atmosphere very much. Once a plume rises above the
  exobase, like a large Pele-type plume, it will produce a
  canopy-shock similar to that on the night-side.  However, whereas at
  night the gas falls down and hits the surface, during the day it
  will encounter the atmosphere, and a re-entry shock develops. This
  will heat the atmosphere up to levels similar to or higher than that seen
  at the vent. The resulting high temperature will lead to excess frost
  sublimation, which gets entrained in the plume flow.
Due to the high pressure created by the high temperature, material will be pushed away, which
 actually results in a decrease in the column density at the intersection between
  the canopy and the atmosphere (i.e., where at night the red ring around Pele
  was created). Some of the material falling down onto the atmosphere
  creating the re-entry shock, will ``bounce'' once, or perhaps
  multiple times, back up and outwards, forming
re-entry shocks everytime when falling down onto the atmosphere
(compare, e.g., the
bounces on the atmosphere seen during the Comet Shoemaker-Levy 9
impact; Nicholson et al., 1995). 
This expands the area of the plume's interaction with the atmosphere by factors of 2--3. The bounces and outward expansion are more
pronunced at higher 
surface temperatures and atmospheric densities. 
Because ultimately the temperature of the
surface frost will maintain the hydrostatic atmosphere in vapor
pressure equilibrium, the total mass in the lower atmosphere may
not change much, but large amounts of plume
material may displace the originally ``sublimed'' gas. The total mass
of material in the plume area will be enhanced, though it is not a
simple addition of the sublimated atmosphere and the plume material
ejected in the absence of a sublimation atmosphere (only 60--75\% of
such a nightside plume is added to the sublimation atmosphere).}

\deleted{In addition to the canopy-shock, a hot
(compared to the background) re-entry
shock is formed due to the interaction of the falling plume gas and
the sublimation atmosphere. The plume-sourced gas is split into two
parts: an inner part that flows back towards the vent, and an outer
part that ``bounces'' multiple times back up and outwards, forming
re-entry shocks when coming down, and 
expanding the area of the plume by factors of 2--3 or more (compare the
bounces on the atmosphere seen during the Comet Shoemaker-Levy 9
impact; Nicholson et al., 1995). The bounces and outward expansion are more
pronunced at higher 
surface temperatures and atmospheric densities. At low surface
temperatures, the SO$_2$ gas quickly condenses when reaching the
surface. As the surface temperature increases, the SO$_2$ frost below
the falling gas sublimates and gets entrained in the plume
flow. The average column density of a day-side Pele plume is
almost an order of magnitude larger than at night, and, as mentioned
above, covers a much larger area. At distances $\gtrsim$150 km the
column density matches that of the background atmosphere.
The authors also simulated a Prometheus-type plume on the day side. This plume is not
sourced from within Io, but most likely caused by the interaction of
molten lava with a thick SO$_2$ frost layer; this explains the 60-km
shift in location between the Voyager and Galileo measurements (Kiefer
et al., 2000). This plume is much smaller than Pele, with a vent
velocity  near 500 m/s, and temperature less than 400 K (McEwen and
Soderblom, 1983). Although this plume is much weaker than Pele and
reaches a height of only 120 km, the plume morphologies are very
similar, and the column densities are also nearly the same, except
right above the Prometheus vent they are roughly a factor of 2 smaller
than for Pele.}

\subsubsection{Comparing ALMA Data with Atmospheric Simulations}\label{sec5d}

\added{\paragraph{Sublimation atmosphere:} During eclipse ingress in March the SO$_2$ flux density decreased
exponentially, caused by a decrease in the volume of SO$_2$
molecules (assuming a hydrostatic atmosphere we showed that the column
density and temperature did not change much; only the fractional area
decreased). With such a tenuous atmosphere, one would expect the
surface temperature to drop instantaneously when entering eclipse, as
shown to be true by Tsang et al (2016). Given a diffusion time of 70
sec (Section~\ref{sec3a.1}; de Pater et al., 2002), the SO$_2$ molecules are expected to
rapidly condense onto the surface, which means that the number density
of molecules just above the surface decreases, resulting in a downward
motion of gas above it. Moore et al. (2009) show that changes occur
primarily in the bottom 10--20 km. They further show that even a small
amount of non-condensible gases will form a diffusion layer near the
surface. Once this layer is several mean-free path-lengths thick, it will
prevent or at least slow down further collapse of the SO$_2$
atmosphere. They predict this to happen after about 20 min. They also
predict that in this case the gas column density and the atmospheric temperature
remain essentially the same. In their
calculations they assumed, though, that SO is non-condensible, while
our data show that SO in essence rapidly condenses through
self-reactions on the surface (the flux density or volume of SO molecules
decreases linearly at a rate of 15 mJy/min). However, since our
observations show essentially no change in column density and
temperature, and some SO$_2$ gas is always present, even when volcanic
activity is low (Fig.~\ref{fig:Sep11Frames}), atmospheric collapse may
indeed be retarded by a diffusive layer of non-condensible gases near
the surface. We cannot exclude the possibility of SO$_2$ emissions due
to stealth volcanism, however.

During eclipse egress in September both the SO$_2$ and SO emissions
increase linearly, though SO is clearly delayed compared to SO$_2$,
which we attributed to formation through photochemistry (Section~\ref{sec5a.2}). The
atmosphere is restored within about 10 min after re-emerging in
sunlight. This suggests that the surface heats up essentially
instantaneously, causing SO$_2$-ice to start to sublime
immediately. This is very different from the calculations by Moore et
al. (2009), who show the atmosphere to reform on a much ($\gtrsim 3\times$) longer
timescale. }

\added{\paragraph{Volcanic plumes:}} The beamsize in our data is $\sim$1200 km, which is similar to the
plume extent of Zhang et al.'s (2003) simulated Pele-type plume at night\deleted{ and the total
Prometheus-type plume on the day side}. A dayside Pele-type plume, though,
when including the effect of bounces, would be resolved in our data. On 2 September, when Io was
in eclipse after having been $\sim$2 hours in the dark, the volcanic
plume over P207 was quite bright, and the line profile very
asymmetric (Fig.~\ref{fig:lineprofilesS}) with a strong blue-shifted
component, indicative of material moving in our direction. Such a line
profile would be expected from Zhang et al.'s (2003) models, as shown
by Moullet et al. (2008), who calculated line profiles based upon
these models at
different locations on Io's disk. Depending on the exact geometry of
the plume, they find kinks or shoulders in the line profile, not unlike
what we observed for P207. The blue-shifted shoulder is caused by the
umbrella-shaped plume material, seen on the limb, moving in our
direction. The model also predicts velocities $v_r$ of order 700 m/s,
which agrees well with the wings in our disk-integrated line profiles,
which are caused by the plumes
(Figs.~\ref{fig:MarchFrames},\ref{fig:Sep2Frames}). 

Within 1--2 minutes after emerging in sunlight, the P207 plume
increased in intensity, and continued to increase for the next several
minutes (Fig.~\ref{fig:Sep2SO2}).  During this period, the plume transitions from a night-side
plume to a day-side plume, when SO$_2$ sublimation from SO$_2$ frost
becomes important.
\added{McDoniel et al. (2017) show calculations of a
  plume transitioning from the night to the dayside, and back into the
  night, a process that takes almost a full Io day (42.5 hrs). The
  ALMA observations, in contrast, show a very accelerated process
  since eclipse egress only takes a few minutes. During these few
  minutes, SO$_2$ frost starts to sublime and the atmosphere reforms,
  while the volcano continues to eject gases.}
The plume starts to interact with the \deleted{subliming}
atmosphere \added{while it is forming. A re-entry shock forms where
  the plume material hits the atmosphere.}  \deleted{falling
  down onto the atmosphere, and in fact accelerates SO$_2$ sublimation where the plume
material falls down and locally heats the subliming atmosphere
through formation of a re-entry shock.} \added{The resulting high
temperature (Section~\ref{sec5c})
accelerates SO$_2$ sublimation, which gets entrained in the plume
flow, causing} the plume area to
grow. \deleted{in particular when multiple bounces are involved.}
\added{Hence, the observed} \deleted{We indeed do
see a} brightening and expansion of the SO$_2$ emissions 
near volcanic vents, i.e., near regions where we see some (though sometimes faint) SO$_2$
emissions during eclipse is consistent with
McDoniel et al.'s simulations. It may also cause the post-eclipse
brightening effect we see about 10 min after eclipse egress, where the
sudden change from night to day and the interaction of the plume
with the reforming atmosphere may lead to a temporary ``excess'' in
SO$_2$ emissions, likely due to an altitude-dependent temporary increase in atmospheric
temperature.

The authors further show that the average
column density over the vent at night is $\sim$10$^{16}$
cm$^{-2}$, and that during the day the column density over the plume matches
that over the dayside hemisphere at distances $\gtrsim$ 150 km. This
essentially agrees with our observations, where column densities
over the plume and background atmosphere on the dayside are very
similar.
The differences in brightness we see between the day- and night-
(eclipse) side, both disk-averaged and over volcanoes, are mostly
explained by changes in the fractional area covered by the gas, but
columns of gas over these areas are very similar. Given our relatively
low spatial resolution this may well be consistent with the models.

The temperature that best matches our line profiles,
$\sim$220--320 K, can be explained qualitatively by the various
temperatures expected along the line-of-sight through the model, which
varies from $\sim$50 K above the vent up to 300--400 K at the canopy
shock. For comparison, when Moullet et al. (2008) parameterized the Zhang et al. (2003)
night-side plumes at a location $\sim$40$^\circ$ away from disk center, they found that the
models could be mimicked well with an isothermal temperature of $\sim$190
K. \added{However, given how complex the plume--atmosphere interaction
  is (McDoniel et al., 2017), we do not think that the atmosphere can
  be modeled correctly using a simple isostatic atmosphere.}

In March we detected vigorous eruptions at Karei and Daedalus
Paterae. At Karei Patera the fractional coverage in sunlight was
roughly 3 times larger than in eclipse, with an atmospheric
temperature of 270$\pm$50 K both in sunlight and in eclipse. As
mentioned before, the entire profile was redshifted by 60 m/s, while
in eclipse the profile was slightly skewed, peaking more at the blue side of
the spectrum. The overall shift towards the red is indicative of plume
material falling back down onto the surface, away from us; since the umbrella-shaped
plume is much larger in extent than the rising column of gas, this can
qualitatively explain the line profiles.

Daedalus Patera, in contrast, 
shows a strong redshifted shoulder, somewhat similar to the blue
shifted shoulder for P207 in September. The entire profile was
slightly blueshifted, presumably because the umbrella-shaped plume material from a
volcanic ejection near the limb has a large component of material
moving towards us (i.e., similar to the material that explains the blue-shifted
wing of the line for P207). The red-shifted wing, though, shows that a large
component of plume material is also moving away from us.

When Moullet
et al. (2008) modeled the Zhang et al. plumes for comparison with
their radio data, they did not see such red-shifted shoulders in the
models. This, together with our observations of line profiles that are
very asymmetric, in particular in-eclipse, shows that the volcanic
eruptions are much more complex than the Zhang et al. (2003) \added{and
McDoniel et al. (2017)} 
models predict. This is not too surprising; volcanic eruptions are
likely not axisymmetric, and may fluctuate in ejection speed, direction, and gas content
on timescales much shorter than we can capture in our
observations. Yet, it is re-assuring that our observations do
qualitatively match many features in the model.

\section{Conclusions}\label{sec6}

We used ALMA in Band 7 (880 $\mu$m) to observe Io in five SO$_2$, two SO, and
one KCl transitions when it went from sunlight into eclipse (20 March 2018),
and from eclipse into sunlight (2 and 11 September 2018). 
We summarize the main findings as follows:

$\bullet$ The disk-averaged brightness temperature at 0.9 mm is
93.6$\pm$5.3 K (including calibration uncertainties). The observed
difference of $\sim$3 K between all Io-in-sunlight and in-eclipse
maps\added{, together with the 22 K drop in temperature at 19 $\mu$m
(Tsang et al., 2016),} 
suggests
that Io's surface \replaced{is overlain with}{is composed of} a thin low-thermal-inertia (50 J
m$^{-2}$ K$^{-1}$ s$^{-1/2}$) layer, overlying a more compact
high-thermal-inertia (320 J m$^{-2}$ K$^{-1}$ s$^{-1/2}$) layer,
indicative of a thin ($\lesssim$few mm) layer of dust or fine-grained
volcanic deposits overlying more compact layers of rock and/or
coarse-grained/sintered ice.  

$\bullet$ The SO$_2$ and SO disk-integrated flux
densities are typically about 2--3 times higher on Io-in-sunlight
than in-eclipse, indicative of a 30--50\% volcanic contribution to the 
emissions. 

$\bullet$ During eclipse ingress, the SO$_2$ flux density dropped
exponentially, with the 346.652 GHz transition (strongest line
intensity) faster and more 
than the other transitions. Following eclipse egress, the
SO$_2$ flux densities increased linearly, with the 346.652 GHz
transition faster than the others. 

$\bullet$ Eclipse egress observations show that the atmosphere is
re-instated on a timescale of 10 minutes,
consistent with the interpretation of the post-eclipse brightening
effect reported for observations of Io's
surface reflectivity. An atmospheric post-eclipse brightening was
seen in several SO$_2$ transitions, where the flux density was up to
$\sim$20\% higher 10 min after re-emerging in sunlight compared to 1/2 hour later.

$\bullet$ We attribute the variations in emissions and differences
between line transitions during eclipse ingress and
egress, as well as the atmospheric post-eclipse brightening effect to 
altitude-dependent changes in temperature, likely caused in/by
volcanic plumes and their interacton with the atmosphere, such as
simulated by \replaced{Zhang et al. (2003)}{McDoniel et al. (2017)}.

$\bullet$ The SO flux density dropped/increased linearly after
entering/re-emerging from eclipse, in both cases clearly delayed compared to
SO$_2$. This provides confirmation that SO may be rapidly removed
through reactions with Io's surface once in eclipse, and that
photolysis of SO$_2$ is a major source of SO. 

$\bullet$ Spectral image data cubes reveal bright volcanic
plumes on 20 March and 2 September; no plumes were detected
on 11 September. Plumes on the limb create high-velocity  wings in
the disk-integrated line profiles (at $\gtrsim$600
kHz, or $\gtrsim$500 m/s). Such high velocities match those predicted
in \deleted{Zhang et al.'s (2003)} plume simulations \added{by Zhang
  et al. (2003) and McDoniel et al. (2017)}.

$\bullet$ In addition to the few obvious
volcanic plumes in our spectral image data-cubes, 
the low level SO$_2$ \deleted{and SO} emissions
present during eclipse may be sourced by stealth volcanic plumes
\added{or be evidence of a layer of non-condensible gases 
  preventing complete collapse of SO$_2$, as modeled by Moore et al. (2009)}.

$\bullet$ Based upon hydrostatic model calculations, typical
disk-integrated SO$_2$ column densities and temperatures are
$N_t \approx (1.5 \pm 0.3) \times 10^{16}$ cm$^{-2}$ and
$T_t \approx 220-320$ K both for Io-in-sunlight and in-eclipse. SO
column densities are roughly a factor of 5--10 lower. The main
differences between in-sunlight and in-eclipse flux densities appear
to be caused by a factor of 2--3 smaller fractional coverage
in-eclipse (i.e., down from 30--35\% SO$_2$ and $\sim$12\% SO in-sunlight).

$\bullet$ The active volcanoes on 20 March and 2 September show
similar
SO$_2$ column densities and temperatures
as for the disk-integrated profiles, but with a very high fractional
coverage ($\sim$50--60\% in sunlight, vs 30-35\% disk-averaged). This seems
consistent with \replaced{Zhang et al.'s (2003)}{McDoniel et al.'s
  (2017)} simulations, where the column
densities blend into the background at distances
\replaced{$\gtrsim$300}{over a few 100} km from
the volcanic vent.

$\bullet$ Line profiles of in-eclipse data are very asymmetric, both
for disk-integrated profiles and individual volcanoes. Some volcanoes
show red-shifted, and others blue-shifted shoulders both in-sunlight
and in-eclipse. Sometimes the entire profile is slightly red- or
blue-shifted. The line profiles must be strongly affected by intrinsic
properties of volcanic
plumes (e.g., ejection speed, direction, density, and variations
therein), in addition to their viewing geometry.

$\bullet$ The data are suggestive of a 300--400 m/s horizontal prograde wind on 11
September, when no volcanic activity was reported; however such a wind
is not supported by 
disk-integrated line profiles. \added{No zonal winds were detected on
  20 March and 2 September, when volcanic plumes were seen.} 

$\bullet$ KCl gas has only been detected on 20 March, sourced
\added{mainly} from near 
Ulgen Patera\deleted{, near Dazhbog Patera, and near Bo\"osaule
Montes}. No SO or
SO$_2$ gas was detected at this location. Hence the magma in the chambers that power volcanoes must
have different melt compositions, and/or the magma has access to different surface/subsurface volatile reservoirs.

$\bullet$ Our data can be qualitatively explained by \deleted{Zhang et
al.'s (2003) plume simulations} \added{the night-side plume simulations of Zhang et
al. (2003) and day-side simulations by McDoniel et al. (2017)}, although it is also clear that the data
are much more complex than the models can capture.

\vskip 0.12truein

Our observations begin to clarify the role of volcanism in forming
Io's atmosphere. However, many questions still remain, including,
e.g., Io's overall atmospheric temperature profile, in particular in
the first 10-20 km above the surface; longitudinal variations in column
density; winds; volcanic sources; magma composition. Although it is
clear that low-level emissions are present during eclipse, we do not
yet understand the cause of these: perhaps stealth
volcanism, or a layer of non-condensible gases preventing complete
collapse of the SO$_2$ atmosphere. To further address these questions we plan to obtain ALMA data
at a higher spatial resolution when the satellite is at eastern and
western  elongation. Future work will also need to include realistic plume models in addition to
the hydrostatic models employed here.

\section*{Acknowledgements}

We are grateful for in-depth reviews by David Goldstein and one
  anonymous referee, which helped improve the manuscript substantially.
This paper makes use of ALMA data  ADS/JAO.ALMA\#2017.1.00670.S. ALMA is a
partnership of ESO (representing its member states), NSF (USA) and
NINS (Japan), together with NRC (Canada), MOST and ASIAA (Taiwan), and
KASI (Republic of Korea), in cooperation with the Republic of
Chile. The Joint ALMA Observatory is operated by ESO, AUI/NRAO, and
NAOJ. The National Radio Astronomy Observatory is a facility of the
National Science Foundation operated under cooperative agreement by
Associated Universities, Inc. The data can be downloaded from the ALMA Archive. 
This research was supported by the National Science Foundation, NSF
grant AST-1313485 to UC Berkeley. PMR acknowledges support from ANID basal AFB170002

\section{References}

Ballester, G. E., M. A. McGrath, D. F. Stobel, X. Zhu, P. D. Feldman,
and H. W. Moos, Detection of the SO$_2$ atmosphere on Io with the Hubble
Space Telescope, Icarus 111, 2-17, 1994.

Bellucci, G., Aversa, E.D., Formisano, V., et al., 2004. Cassini/VIMS
observation of an Io post-eclipse brightening event. Icarus, 172, 141-148.

Binder, A.P., Cruikshank, D.P., 1964. Evidence for an atmosphere on
Io. Icarus 3, 299-305.

Cornwell, T. J., and E. B. Fomalont 1999. Self-calibration. In
Synthesis Imaging in Radio Astronomy II (G. B. Taylor, C. L. Carilli,
and R. A. Perley, Eds.), pp. 187-199, ASP Conf. Series. Astron. Soc. of the Pacific, San Francisco.

Cruikshank, D.P., Emery, J.P., Korney, K.A., Bellucci, G., Aversa, E.,
2010. Eclipse reappearances of Io: Time resolved spectroscopy
(1.9--4.2 $\mu$m). Icarus, 205, 516-527.

de Kleer, K., I. de Pater, 2016. Time Variability of Io's Volcanic Activity from Near-IR Adaptive Optics
Observations on 100 Nights in 2013-2015. Icarus, 280, 378-404. 

de Kleer, K., de Pater, I., Molter, E., Banks, E, Davies, A.,
Alvarez, C., Campbell, R., et al., 2019. Io's volcanic activity from
Time-domain Adaptive Optics Observations: 2013--2018.  Astron. J.,
 158, 129 (14pp).  https://doi.org/10.3847/1538-3881/ab2380

de Kleer, K., Butler, B., de Pater, I., Gurwell, M., Moullet, A.,
Trumbo, S., Spencer, J., 2020. Thermal properties of Ganymede’s
surface from millimeter and infrared emission. PSJ ??, in prep.

de Pater, I., H.G. Roe, J.R. Graham, D.F. Strobel, and P. Bernath,
2002.  Detection of the Forbidden SO $a^1 \Delta \rightarrow X^3
\Sigma^-$ Rovibronic Transition on Io at 1.7 $\mu$m. {\it Icarus Note} {\bf
156}, 296-301.

de Pater, I., F. Marchis, B.A. Macintosh, H.G. Roe, D. Le
Mignant, J.R. Graham, and A.G. Davies, 2004. Keck AO observations of
Io in and out of eclipse.  Icarus, 169, 250-263.

de Pater, I., C. Laver, F. Marchis, H.G. Roe, and B.A. Macintosh, 2007. 
Spatially Resolved Observations of the Forbidden SO $a^1 \Delta \rightarrow X^3
\Sigma^-$ Rovibronic Transition on Io during an Eclipse. Icarus,
191, 172-182.

de Pater, I., Sault, R.J., Moeckel, C., Moullet, A., Wong, M.H.,
Goullaud,C., DeBoer,D., Butler, B., Bjoraker, B., \'Ad\'amkovics, M.,
Cosentino, R., Donnelly, P.T., Fletcher, L.N., Kasaba, Y., Orton, G.,
Rogers, J., Sinclair, J., Villard, E., 2019. First ALMA millimeter wavelength maps of Jupiter,
with a multi-wavelength study of convection. {\it Astron. J.}, {\bf
158}, 139 (17 pp). https://doi.org/10.3847/1538-3881/ab3643

de Pater, I., de Kleer, K., \'Ad\'amkovics, M., 2020. High Spatial
and Spectral Resolution Observations of the Forbidden 1.707 $\mu$m
Rovibronic SO Emissions on Io. {\it Planetary Science Journal},
submitted.

Fanale, F.P., Banerdt, W.B., Cruikshank, D.P., 1981. IO: Could SO$_2$
condensation/sublimation cause the sometimes reported post-eclipse
brightening? GRL, 8, 625-628.

Feaga, L.M., McGrath, M., Feldman, P.D., 2009. Io's dayside SO$_2$
atmosphere. Icarus 201, 570-584.

Geissler, P.E., McEwen, A.S., Keszthelyi, L., Lopes-Gautier, R.,
Granahan, J., Simonelli, P., 1999. Global color variations on Io,
Icarus, 140, 265-282.

Geissler, P.E., Smyth, W.H., McEwen, A.S.,  et al.,  2001. Morphology
and time variability of 	Io's visible aurora. J. Geophys. Res.,
106, 26,137-26,146.

\added{Geissler, P.E., McEwen, A.S., Philips, C., Keszthelyi, L., Spencer,
J., 2004. Surface changes on Io during the Galileo mission. Icarus,
169, 29-64.}

Geissler, P., McEwen, A., Porco, C., Strobel, D., Saur, J., Ajello,
J., West, R. 2004.
Cassini observations of Io's visible aurorae. Icarus, 172, 127-140.

\added{Gratiy, S.L., Walker, A.C., Levin, D.A., Goldstein, D.B.,
  Verghese, P.L., Trafton, L.M., Moore, C.H., 2010. Multi-wavelength simulations of atmospheric radiation from Io with a 3-D
spherical-shell backward Monte Carlo radiative transfer model. Icarus,
207, 394-408.}

Hammel, H.B., Goguen, J.D., Sinton, W.M., Cruikshank, D.P.,
1985. Observational Tests for Sulfur AIIotropes on Io. Icarus, 64, 125-132.

\added{Hapke, B., Graham, F., 1989. Spectral properties of condensed phases 
of disulfur monoxide, polysulfur oxide, and irradiated sulfur. Icarus, 
79, 47–55.}


\added{Ingersoll, A.P., Summers, M.E., Schlipf, S.G., 1985. Supersonic meteorology of Io:
Sublimation-driven flow of SO$_2$. Icarus 64, 375-390.}

Jessup, K.L., Spencer, J.R., Yelle, R., 2007. Sulfur volcanism on Io. Icarus 192, 24-40.

Johnson, T. V., Matson, D. L., Blaney, D.L., Veeder, G.J., Davies, A.,
1995. Stealth plumes on Io, Geophys. Res. Lett. 22, 3293-3296.

Kieffer, S.W., Lopes-Gautier, R., McEwen, A., Smythe, W., Keszthelyi, L., Carlson, R., 2000. Prometheus: Io's wandering plume. Science 288,
1204 -1208.

\added{Kumar, S. 1985. The SO$_2$ atmosphere and ionosphere of Io: Ion
  chemistry, atmospheric escape, and models corresponding to the
  Pioneer 10 radio occultation measurements. Icarus, 61, 101-123.}

Lellouch, E., M.J.S. Belton, I. de Pater, S. Gulkis, and T. Encrenaz,
1990, Io's atmosphere from microwave detection of SO$_2$,  Nature,
346, 639-641. 

Lellouch, E., M. Belton, I. de Pater, G. Paubert, S. Gulkis,
and Th. Encrenaz, 1992, The Structure, Stability, and Global Distribution of 
Io's Atmosphere, Icarus, 98, 271-295.

Lellouch, E., 1996. Urey Prize Lecture. Io's Atmosphere: Not Yet Understood.
Icarus, 124, 1-21.

Lellouch, E., Paubert, G., Moses, J. I., Schneider, N. M., and Strobel, D. F., 2003. Volcanically-emitted sodium chloride as a
source for Io's neutral clouds and plasma torus. Nature 421, 45-47.

Lellouch, E., Strobel, D. Belton, M., Paubert, G., Ballester, G., de
Pater, I., 1994, Millimeter wave obervations of Io's atmosphere: new data
and new models, BAAS, 26, 1136.

\added{Lellouch, E., Paubert, G., Moses, J. I., Schneider, N. M., and Strobel, D. F. 2003. Volcanically-emitted sodium chloride as a source for Io's neutral clouds and plasma torus.
  Nature, 421, 45-47.}

Lellouch, E., McGrath, M. A., Jessup, K. L., 2007. Io's atmosphere, in
{\it Io after Galileo: A New View of Jupiter's Volcanic
  Moon}. Eds. Lopes, R.M., \& Spencer, J. R. Springer. ISBN
3-540-34681-3. 

 Lellouch, E., Ali-Dib, M., Jessup, K.-L., Smette, A., K\"aufl, H.-U.,
 Marchis, F., 2015. Detection and characterization of Io's atmosphere
 from high-resolution 4-$\mu$m spectroscopy. Icarus, 253, 99-114.

Macintosh, B., D. Gavel, S.G. Gibbard, C.E. Max, I. de Pater,
A. Ghez, and J. Spencer. 2003. Speckle imaging of volcanic hot spots
on Io with the Keck telescope. Icarus, 165, 137-143.

Marchis, F., D. Le Mignant, F. Chaffee, A.G. Davies, T. Fusco, R.
Pranage, I. de Pater and the Keck Science team, 2005. Keck AO
survey of Io's global volcanic activity between 2 and 5$\mu$m. 
Icarus,  176, 96-122.

\added{McDoniel, W.J., Goldstein, D.B., Varghese, P.L., and Trafton, L.M., 2017. The interaction of Io's plumes and sublimation atmosphere. Icarus, 294, 81-97.}

McEwen, A.S., Soderblom, L.A., 1983. Two classes of volcanic plumes on
Io. Icarus 58, 197-226.

McGrath, M.A., Lellouch, E., Strobel, D.F., Feldman, P.D., Johnson,
R.E., 2004. Satellite atmospheres.  In {\it Jupiter: Planet,
Satellites \& Magnetosphere}, Eds. F. Bagenal, T. E. Dowling, 
  and W. McKinnon,
  pp. 457-483. Cambridge, UK: Cambridge University Press.

 McGrath, M.A., Belton, M.J.S., Spencer, J.R., Sartoretti, P.,
 2000. Spatially resolved spectroscopy of Io's Pele plume and SO$_2$ atmosphere. Icarus 146, 476-493.

 Moore, C., Goldstein, D. B., Varghese, P., Trafton, L. and Stewart,
 B. 2009. 1-D DSMC simulation of Io's atmospheric collapse in
 eclipse. Icarus, 201, 585-597.
 
Morrison, D., Cruikshank, D. P. 1973. Thermal properties of the Galilean satellites.
Icarus, 18, 224-236.
 
Moses, J. I., Nash, D.B., 1991. Phase transformations and the spectral
reflectance of solid sulfur - Can metastable sulfur allotropes exist
on Io?, Icarus 89, 277-304.

Moses, J.I., Zolotov, M.Y., Fegley, B. Jr., 2002. Photochemistry of a
volcanically driven atmosphere on Io: Sulfur and oxygen species from a
Pele-type eruption. Icarus, 156, 76-106.

Moullet, A., Lellouch, E., Moreno, R., Gurwell, M.A., Moore, C., 2008. First disk-
resolved millimeter observations of Io's surface and SO$_2$ atmosphere. Astrophys.
Astron. 482, 279-292. doi:10.1051/0004-6361:20078699.

Moullet, A., Gurwell, M.A., Lellouch, E., Moreno, R.,
2010. Simultaneous mapping of SO$_2$, SO, NaCl in Io's atmosphere with
the Submillimeter Array. Icarus, 208, 353-365.

Moullet, A. et al., 2013. Exploring Io’s atmospheric composition with
APEX: First measurement of 34SO2 and tentative detection of
KCl. Astrophys. J. 776,
32. http://dx.doi.org/10.1088/0004-637X/776/1/32, 9 pp.

Moullet, A., 2015. Exploring the Solar System with ALMA. 2015. {\it Revolution in
Astronomy with ALMA: The third year}. ASP Conference Series, Vol. 499.
Daisuke Iono, Ken'ichi Tatematsu, Al Wootten, Leonardo Testi, eds. Astronomical Society of the Pacific

\added{Moullet, A, Lellouch, E, Gurwell, M, Moreno, R, Black, J, Butler, B. 2015. AAS DPS Meeting \#47,
Abstract 311.31}

Nicholson, P. D., P. J. Gierasch, T. L. Hayward, C. A. McGhee,
J. E. Moersch, S. W. Squyres, J. Van Cleve, K. Matthews,
G. Neugebauer, D. Shupe, A. Weinberger, J. W. Miles, and
B. J. Conrath, 1995. Palomar observations of the R impact of comet
Shoemaker-Levy 9: I. Light curves, Geophys. Res. Lett. 22, 1613–1616.

Pearl, J., Hanel, R., Kunde, V., Maguire, W., Fox, K., Gupta, S., Ponnamperuma, C.,
Raulin, F., 1979. Identification of gaseous SO$_2$ and new upper limits for other
gases on Io. Nature 280, 755-758. doi:10.1038/280755a0.

\added{Rathbun, J.A., Spencer, J.R., Tamppari, L.K., Martin, T.Z., Barnard,
L., Travis, L.D., 2004. Mapping of Io's thermal radiation by the
Galileo photopolarimeter-radiometer (PPR) instrument. Icarus, 169, 127-139.}

Retherford, K.D., Spencer, J.R., Stern, S.A., et al., 2007. Io's
atmospheric response to eclipse: UV aurorae observations. Science 318,
237-241.

Roesler, F.L., Moos, H.W., Oliversen, R.J., Woodward, R.C., Retherford, K.D., Scherb, F., McGrath, M.A., Smyth, W.H., Feldman, P.D.F., Stro bel, D., 1999. Far-ultraviolet imaging spectroscopy of Io's atmosphere
with HST/STIS. Science 283, 353-357.

Saur, J., Strobel, D.F., 2004. Relative contributions of sublimation
and volcanoes to Io's atmosphere inferred from its plasma interaction
during solar eclipse. Icarus, 171, 411-420.

Schenk, P. W., Steudel, R., 1965. New findings in the chemistry of the lower
oxides of sulfur. Angewandte Chem. Intern. Ed. Engl. 4, 402-409.


Sinton, W.M., Kaminsky, C., 1988. Infrared observations of eclipses of
Io, its thermophysical parameters, and the thermal radiation of the
Loki volcano and environs.
Icarus, 75, 207-232.

Spencer, J. R., 1987. The Surfaces of Europa, Ganymede, and Callisto: An
Investigation Using Voyager IRIS Thermal Infrared Spectra,
Ph.D. thesis, University of Arizona.

Spencer, J.R., N.M. Schneider, 1996. Io on the eve of the Galileo
mission. Ann. Rev. Earth and Plan. Sci., 24, 125-190.

Spencer, J. R., L. K. Tamppari, T. Z. Martin, and L. D. Travis,
1999. Temperatures on Europa from Galileo PPR: Nighttime thermal
anomalies, Science 284, 1514–1516, 1999.

Spencer, J.R., Lellouch, E., Richter, M.J., L\'opez-Valverde, M.A., Lea Jessup, K., Greathouse, T.K., Flaud, J.-M., 2005. Mid-infrared detection of large longitudinal asymmetries in Io's SO$_2$ atmosphere. Icarus 176, 283-304.

Spencer, J.R., Stern, S.A., Cheng, A.F., et al., 2007. Io volcanism
seen by New Horizons: A major eruption of the Tvashtar
volcano. Science, 318, 240-243.

\added{Strobel, D.F., Zhu, X., Summers, M.E., 1994. On the vertical structure of Io's
  atmosphere. Icarus 111, 18-30.}

\added{Summers, M. E. 1985. Theoretical studies of Io's
  atmosphere. Ph.D. thesis, California Institute of Technology,
  Pasadena.}

Tsang, C.C.C., Spencer, J.R., Jessup, K.L., 2015. Non-detection of
post-eclipse changes in Io's Jupiter-facing atmosphere: Evidence for
volcanic support? Icarus, 248, 243-253.

Tsang, C.C.C., Spencer, J.R., Lellouch, E., Lopes-Valverde, M.A., Richter, J.J., 2016. The
collapse of Io's primary atmosphere in Jupiter eclipse. JGR 121, 1400-1410.

Wagman, D. D. 1979. Sublimation Pressure and Enthalpy of
S0$_2$. Chem. Thermodynamics Data Center, Nat. Bureau of Standards,
Washington, DC.

Walker, A.C., Gratiy, S.L., Goldstein, D.B., Moore, C.H., Varghese, P.L., Trafton, L.M.,
Levin, D.A., Stewart, B.D., 2010. A comprehensive numerical simulation of Io's
sublimation driven atmosphere. Icarus 207, 409-432.

Walker, A., Moore, C., Goldstein, D., Varghese, P., Trafton, L.
2012. A parametric study of Io's thermophysical surface parameters and
subsequent numerical atmospheric simulations based on the best fit
parameters. Icarus, 220, 225-253. 

Zhang, J., Goldstein, D. B., Varghese, P. L., Gimelshein, N. E.,
Gimelshein, S. F., and Levin, D. A. 2003. Simulation of gas dynamics
and radiation in volcanic plumes of Io. Icarus, 163, 182-187.

\added{Zhang, J., Goldstein, D., Varghese, P., Trafton, L., Moore, C., Miki, K., 2004. Numerical modeling of Ionian volcanic plumes with entrained particulates. Icarus, 172 , 479-502.}



\begin{figure*}
\includegraphics[scale=0.85]{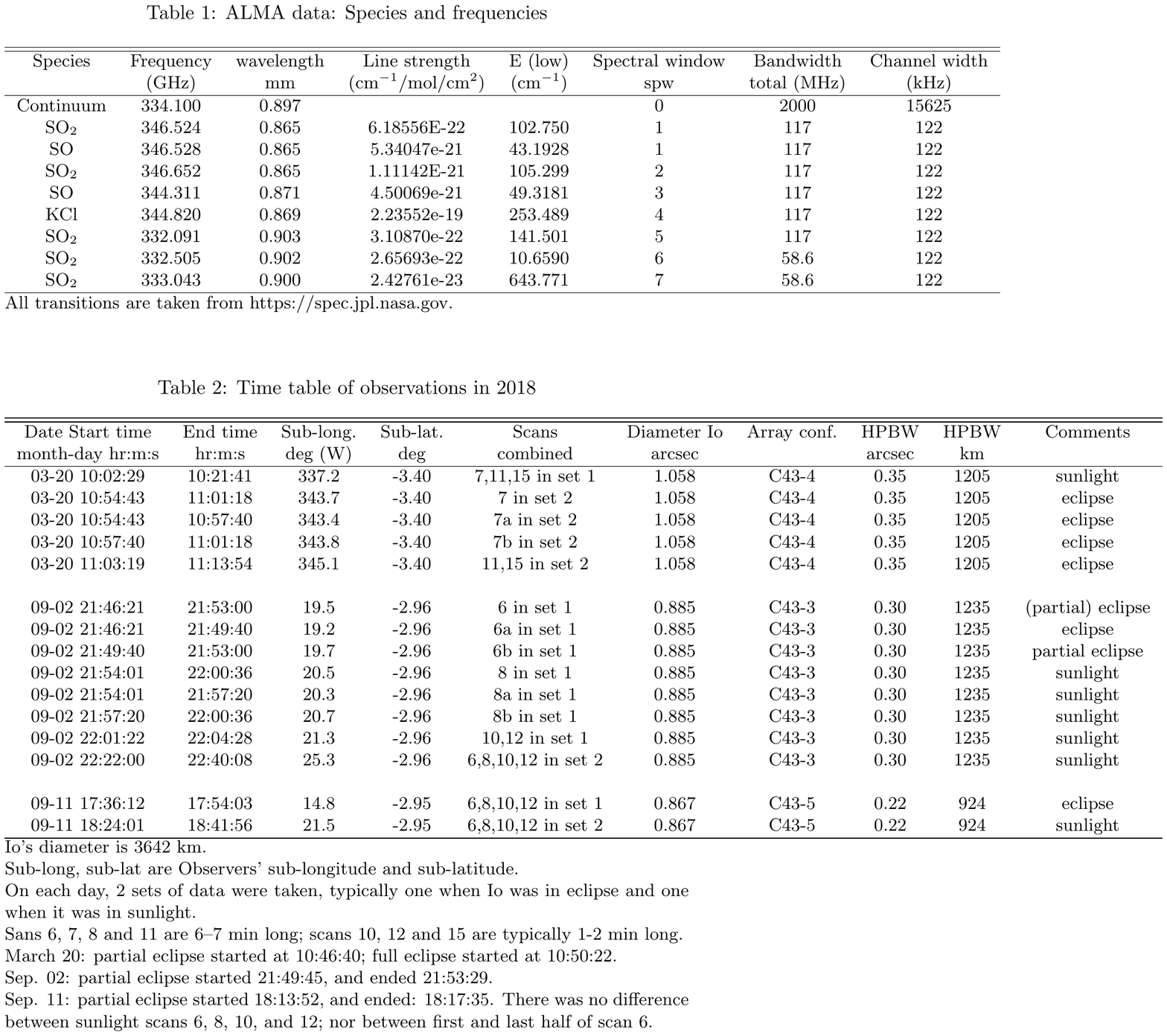}
\end{figure*}

\begin{figure*}
\includegraphics[scale=0.9]{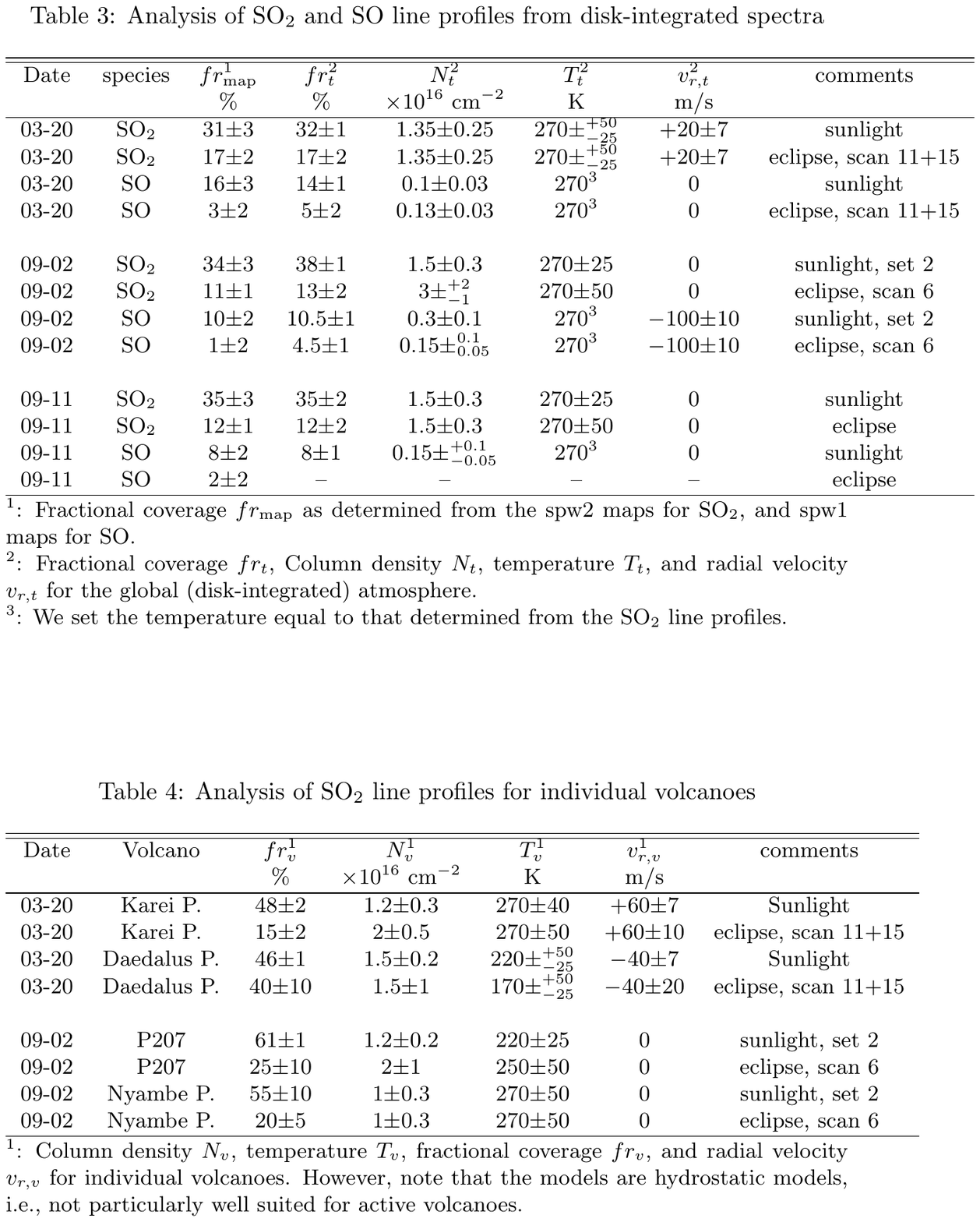}
\end{figure*}

\setcounter{table}{0}
\setcounter{figure}{0}

\begin{figure*}
\includegraphics[scale=0.5]{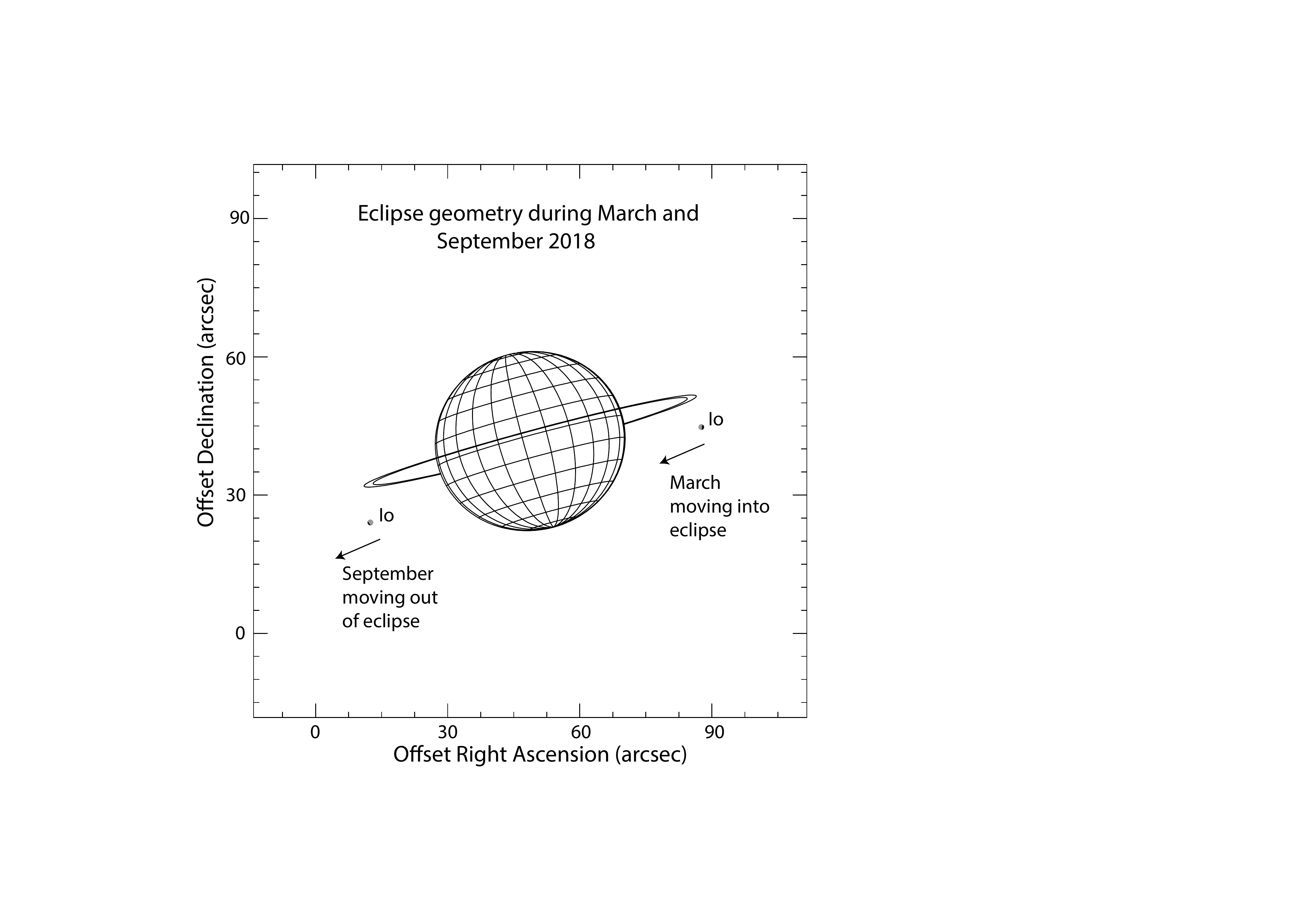}
\caption{The geometries of Io moving into eclipse (March 2018) and
  coming out of eclipse (September 2018). (Adapted from the Planetary Ring Node: http://pds-rings.seti.org/tools/). 
}
\label{fig:geometry}\end{figure*}

\begin{figure*}
\includegraphics[scale=0.5]{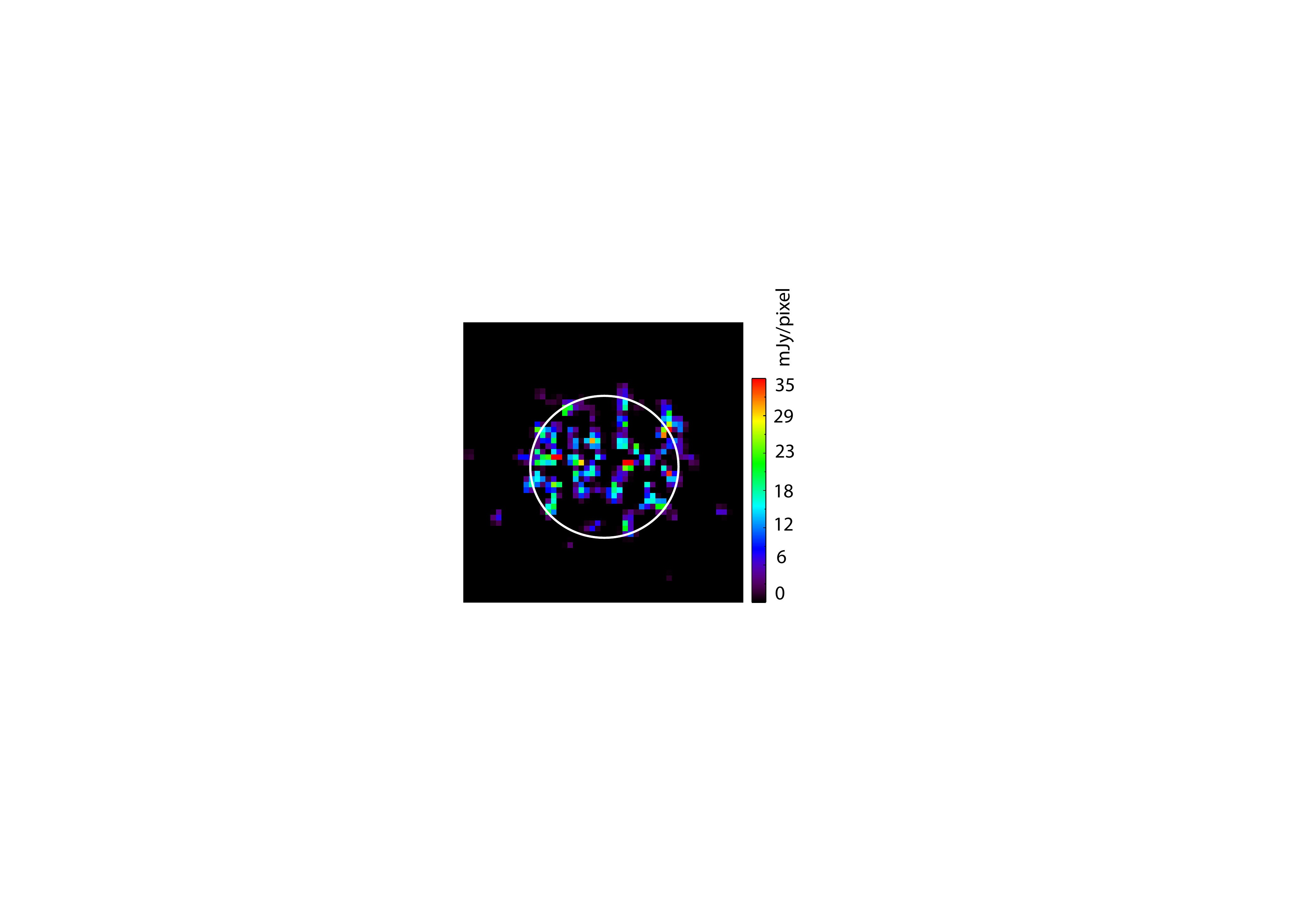}
\caption{This .model map shows the sum of all CLEAN components per
  pixel as obtained from CASA's tCLEAN routine when deconvolving the
  original Io-in-sunlight map at 346.652 GHz. After convolution with
  the HPBW, and restoration to the residual map, this particular
  .model map results in the map displayed in the top left panel of Fig.~\ref{fig:MarchSO2}.
}
\label{fig:cc}\end{figure*}

\begin{figure*}
\includegraphics[scale=0.7]{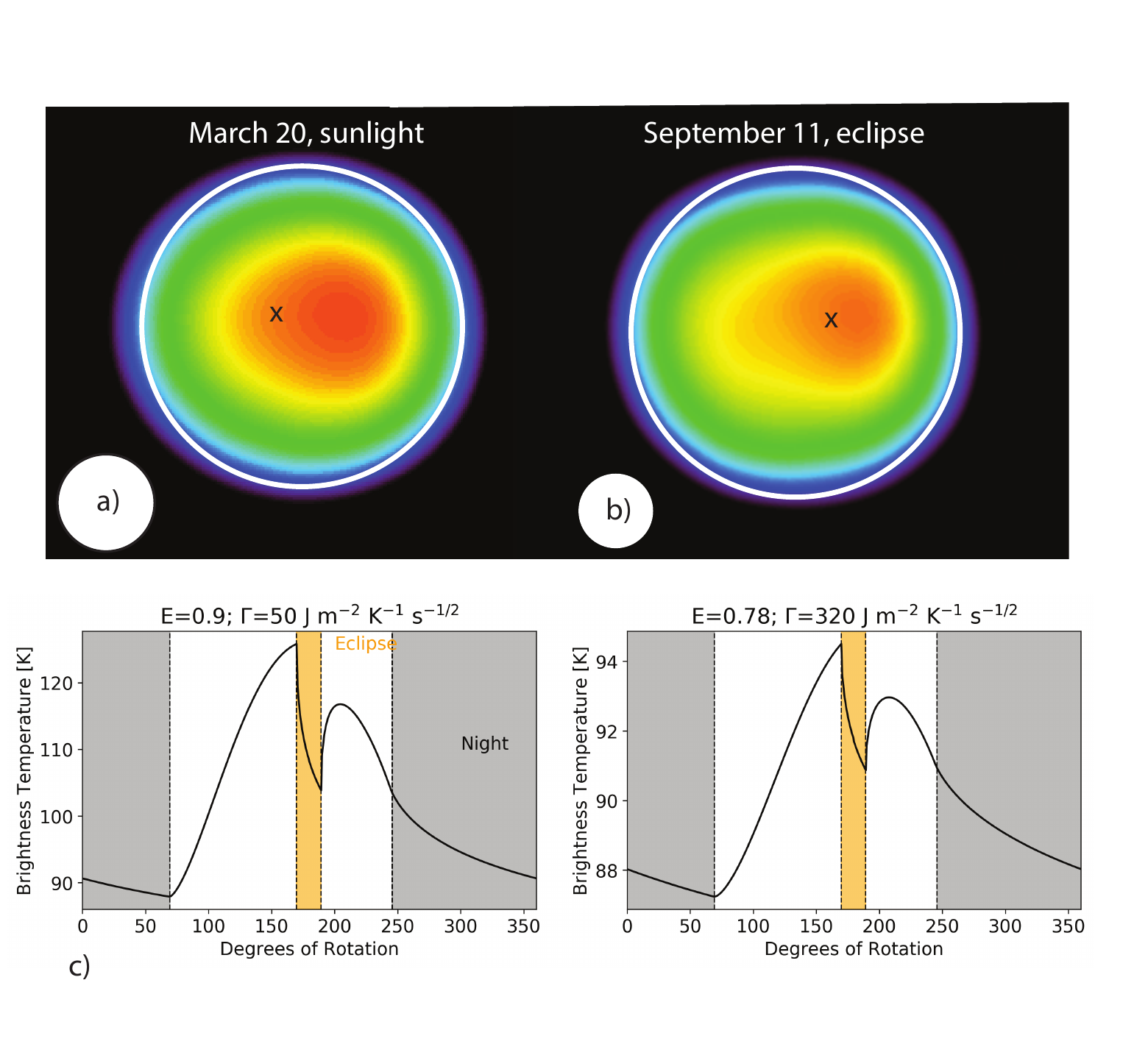}
\caption{Continuum image of Io at 334.1 GHz taken on 20 March 2018 while Io was in
  sunlight (panel a), and on 11 September while Io was in eclipse
  (panel b). Io North is up in these images. \added{The white circle
    shows the approximate size of Io's disk.} The X indicates the
  approximate sub-solar location, and the approximate beam size
  is indicated in the lower left corner. The temperature scale is from 0 to $\sim$90 K, but not quite
  linear to bring out the slight asymmetry in the emission. c) Simple
thermal conduction model at mid-latitudes that can explain the differences in
brightness temperature between the infrared and millimeter
data when entering an eclipse. (see text for details).
}
\label{fig:cont}\end{figure*}

\begin{figure*}
\includegraphics[scale=0.5]{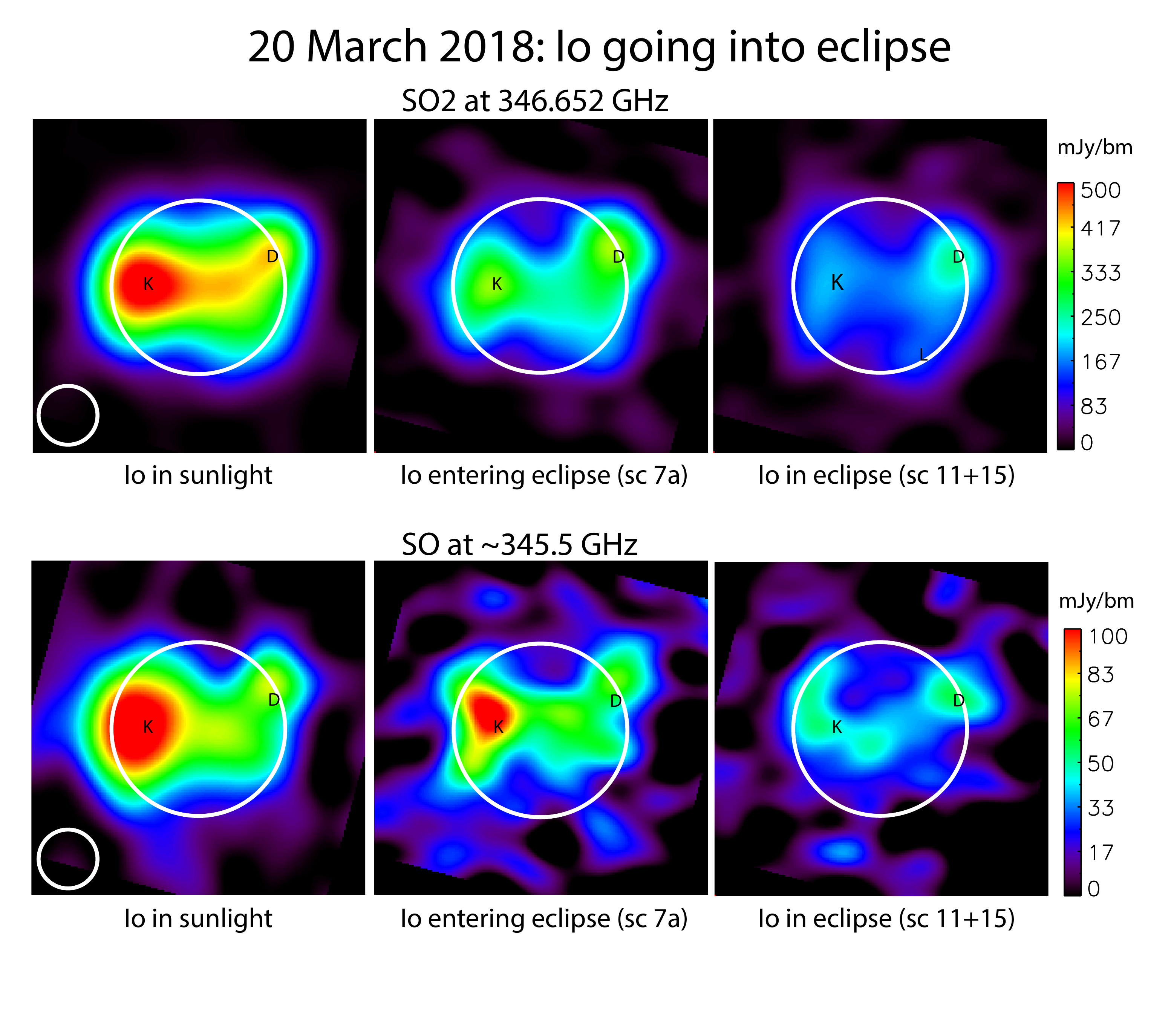}
\caption{Top row: Maps of the spw2 data  of the SO$_2$ distribution on Io-in-sunlight, and
  $\sim$~6 (scan 7a) and $\sim$~15 (scans 11$+$15) min after entering
  eclipse. Bottom row: maps of the averaged spw1 \& spw3 SO data taken at the
  same times as the SO$_2$ maps. All maps were averaged over 0.4km/s
  ($\sim$0.45 MHz). Io North is up in all frames. The large circle shows the outline
  of Io, and the small circle in the lower left shows the
  size of the beam (HPBW). The volcanoes Karei Patera (K), Daedalus
  Patera (D), and North Lerna (L, on one panel only) are indicated.
  }
\label{fig:MarchSO2}\end{figure*}

\begin{figure*}
\includegraphics[scale=0.35]{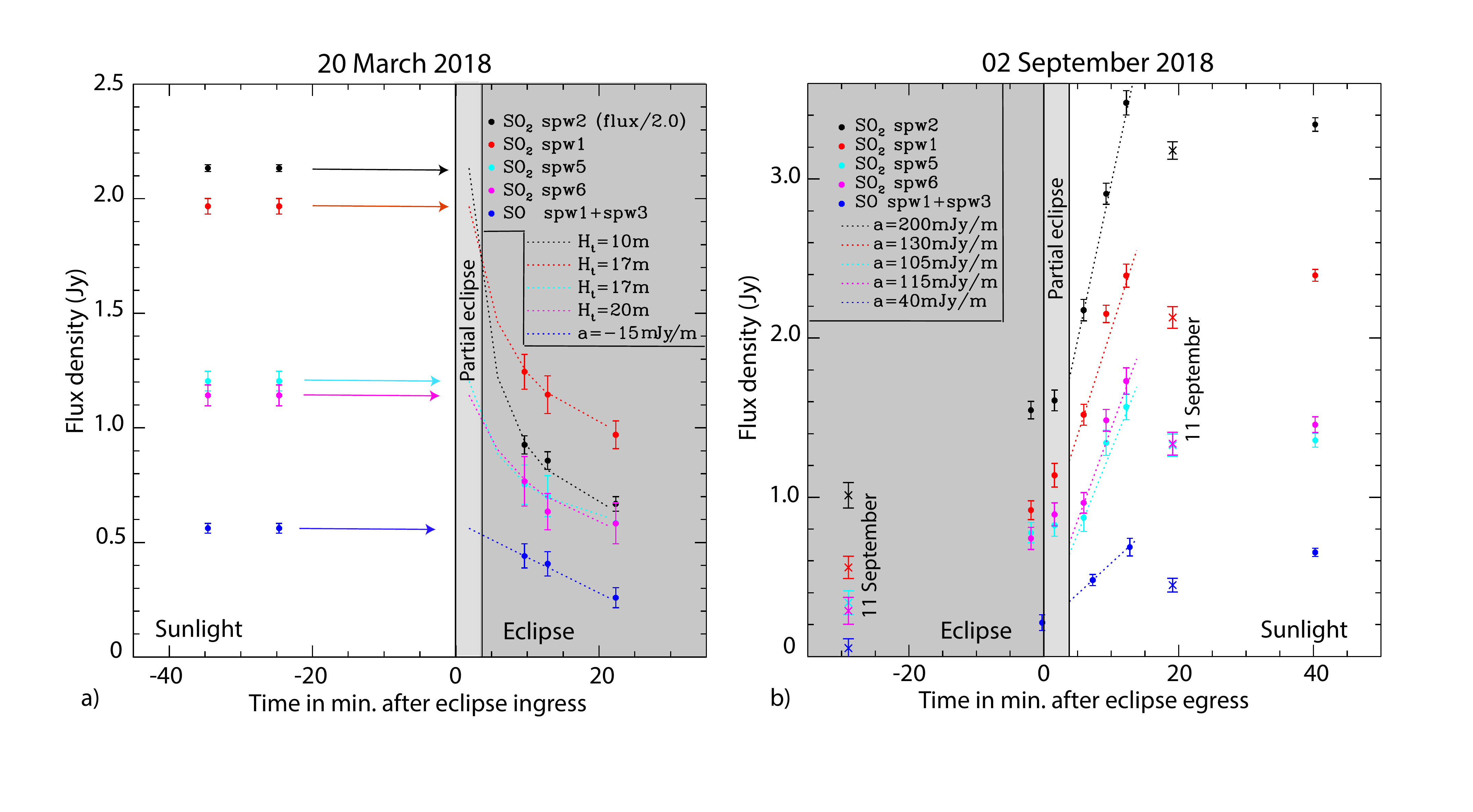}
\caption{Flux densities integrated over individual maps (as in 
  Figs.~\ref{fig:MarchSO2}, \ref{fig:Sep2SO2}, and \ref{fig:Sep11SO2})
  as a function of time (filled circles for March 20 and September 2
 \added{; crosses (x) for Sep. 11}). The colors refer to different
  spectral windows. The data for SO were averaged over spw1 and
  spw3 to increase the SNR. The dotted lines superposed on the data in panel a show the exponential
  decrease \deleted{in the first few minutes} (equ. 1) or
  the linear slope (equ. 2) after entering eclipse,
  whichever is appropriate. In panel b the dotted lines show
  the linear increase after emerging from eclipse on September 2.\deleted{In
  this panel the data for September 11 are shown as open squares,
  separated on the figure by vertical dashed lines. The eclipse data
  are shown at approximately the right time compared to the partial
  eclipse; the in-sunlight data should be at time stamp 69 min, about
  17 min after eclipse egress (Table 2).} All data are
  normalized to a geocentric distance of 5.044 AU.
   }
\label{fig:ec_sun}\end{figure*}

\begin{figure*}
\includegraphics[scale=0.5]{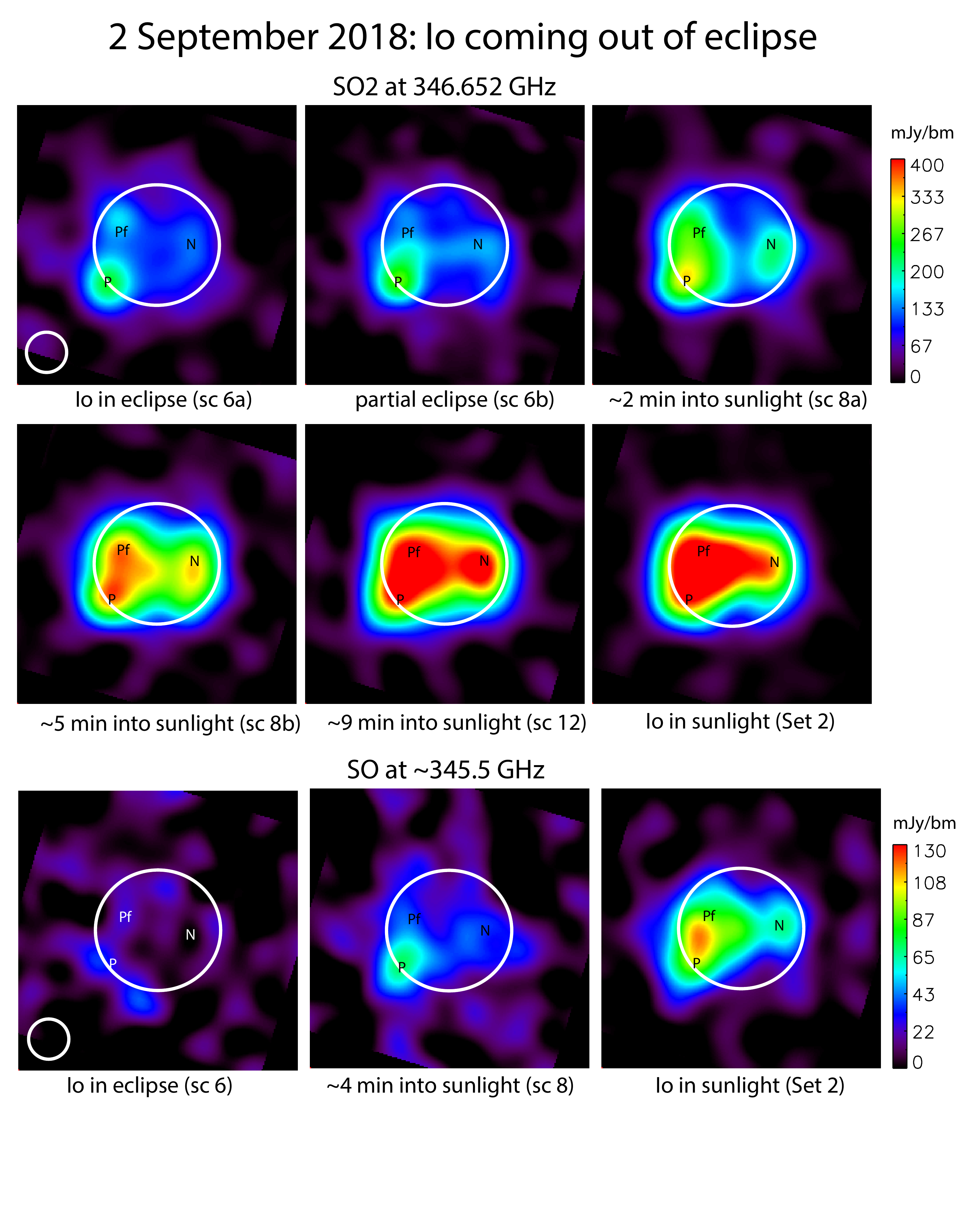}
\caption{Top 2 rows: Maps of the spw2 data of the SO$_2$ distribution on Io in eclipse (scan 6a),
  and emerging into sunlight on 2 Sep. 2018, starting with a partial eclipse (scan
  6b), as indicated. Bottom row: maps of the averaged spw1 \& spw3 SO
  data.  All maps were averaged over 0.4km/s
  ($\sim$0.45 MHz). See Table 2 for exact times of each scan. Io
  North is up.
  The large circle shows the outline
  of Io. The small circle in the lower left shows the
 HPBW. The letters show the positions of several volcanoes:
  P for P207; Pf for
  PFd1691, and N for Nyambe Patera.
  }
\label{fig:Sep2SO2}\end{figure*}

\begin{figure*}
\includegraphics[scale=0.5]{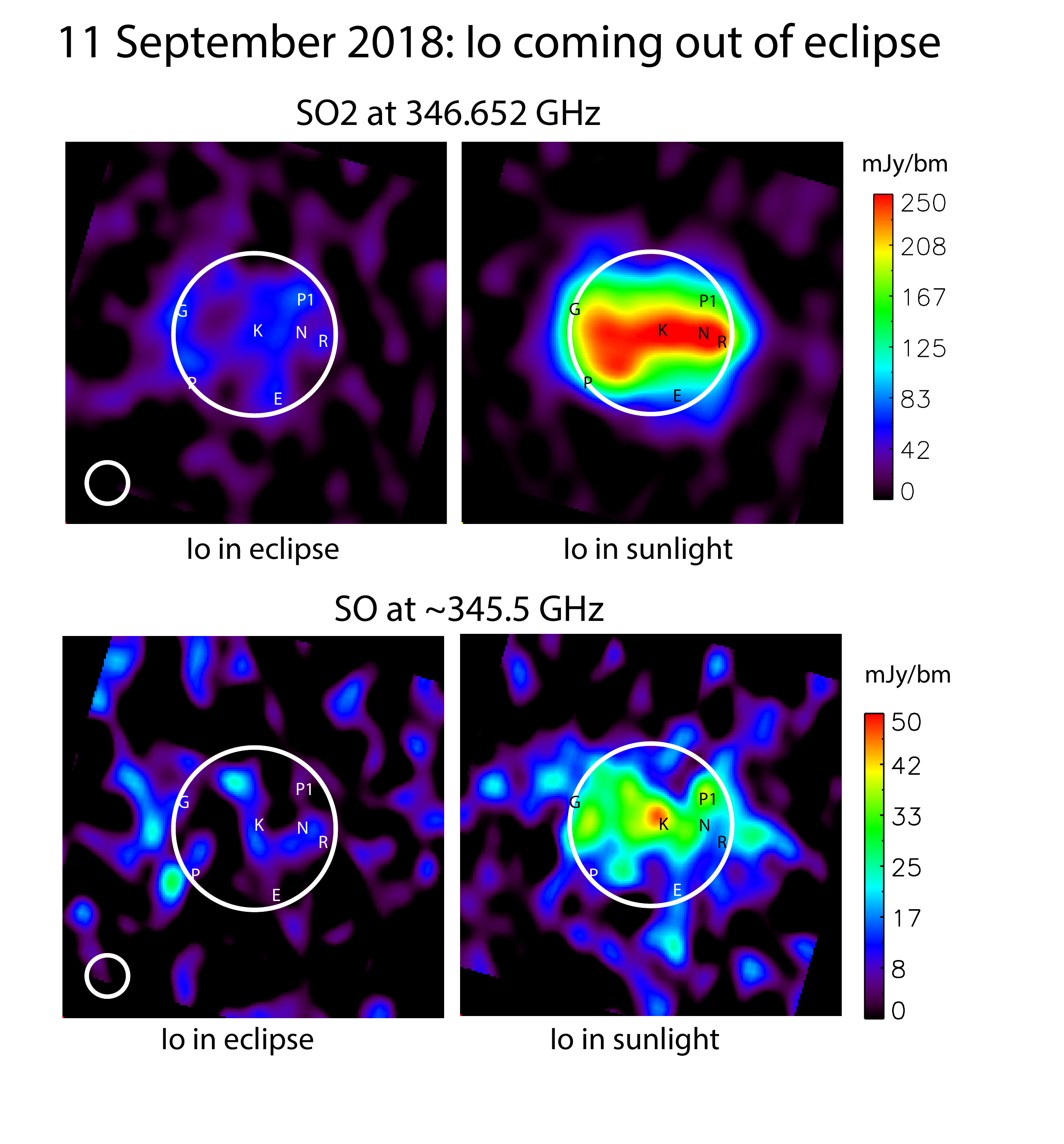}
\caption{Top row: Maps of the spw2 data of the SO$_2$ distribution on Io in eclipse 
  and in sunlight on 11 Sep. 2018. Bottom row: Maps of the SO
  distribution on Io in sunlight, and in eclipse on 11 Sep. 2018. The maps from spw1 and spw3 were
    averaged to increase the SNR. Io
  North is up in these frames. All maps were averaged over 0.4km/s
  ($\sim$0.45 MHz). 
  The large circle shows the outline
  of Io. The small circle in the lower left shows the
 HPBW. Note that the beam is smaller than in figures
 \ref{fig:Sep2SO2} and \ref{fig:MarchSO2}, so that the intensity scale
 on the right shows values that are much smaller than in the other figures.
The letters show the positions of several volcanoes:
 P: P207; G: Gish Bar Patera; K: Karei Patera; N:
  Nyambe Patera; P1: P129; R: Ra Patera, E: Euboea.
  }
\label{fig:Sep11SO2}\end{figure*}

\begin{figure*}
\includegraphics[scale=0.5]{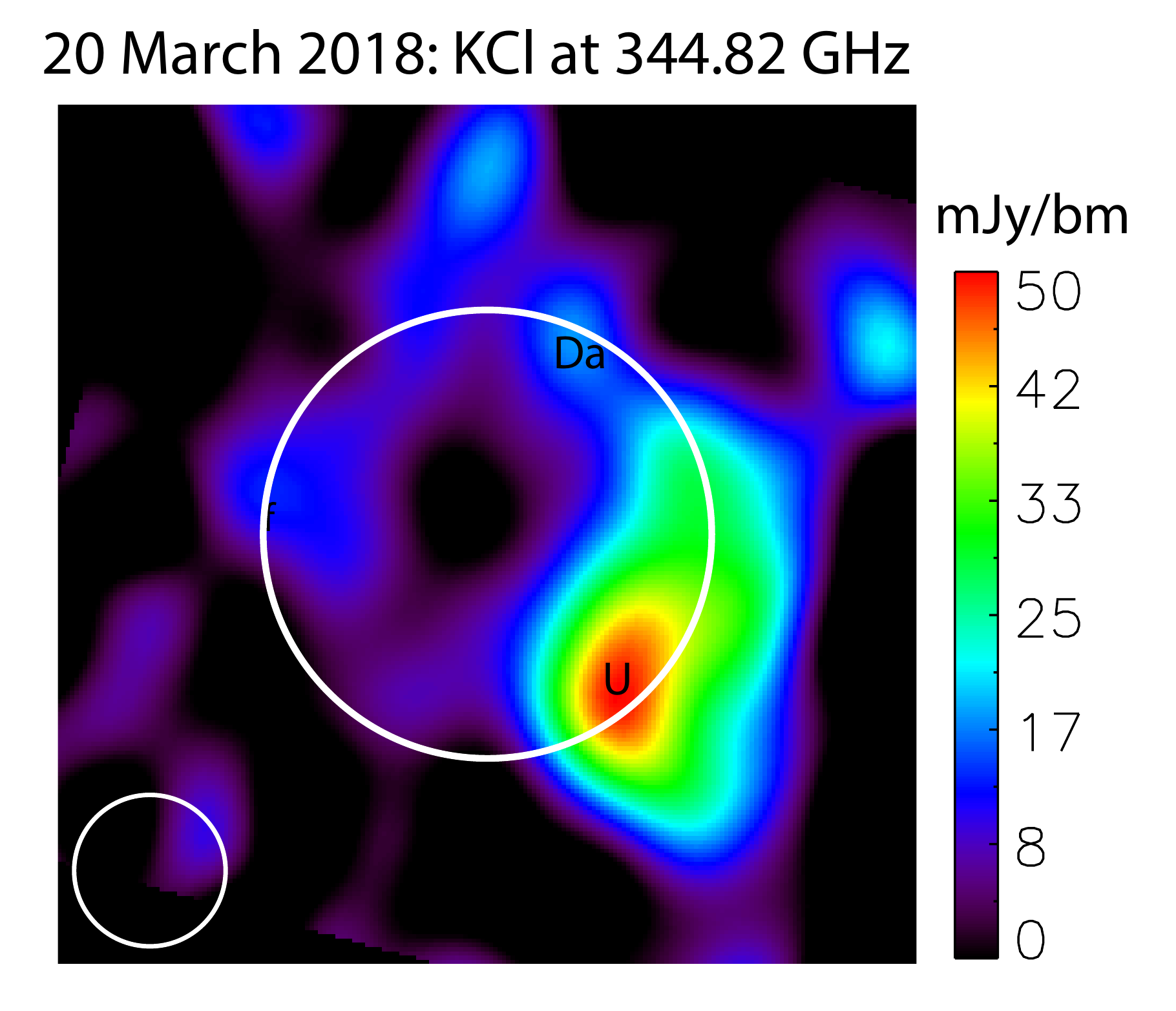}
\caption{Map of the spatial distribution of KCl on 20 March
  2018. The map was averaged over 0.4km/s
  ($\sim$0.45 MHz). Io
  North is up in this frame. This map is from the sunlight data only,
  and is essentially the same as one in which sunlight and eclipse
  data are averaged. The volcanoes indicated on this map are: U: Ulgen Patera; Da: Dazhbog Patera.
  }
\label{fig:MarchKCL}\end{figure*}

\begin{figure*}
\includegraphics[scale=0.48]{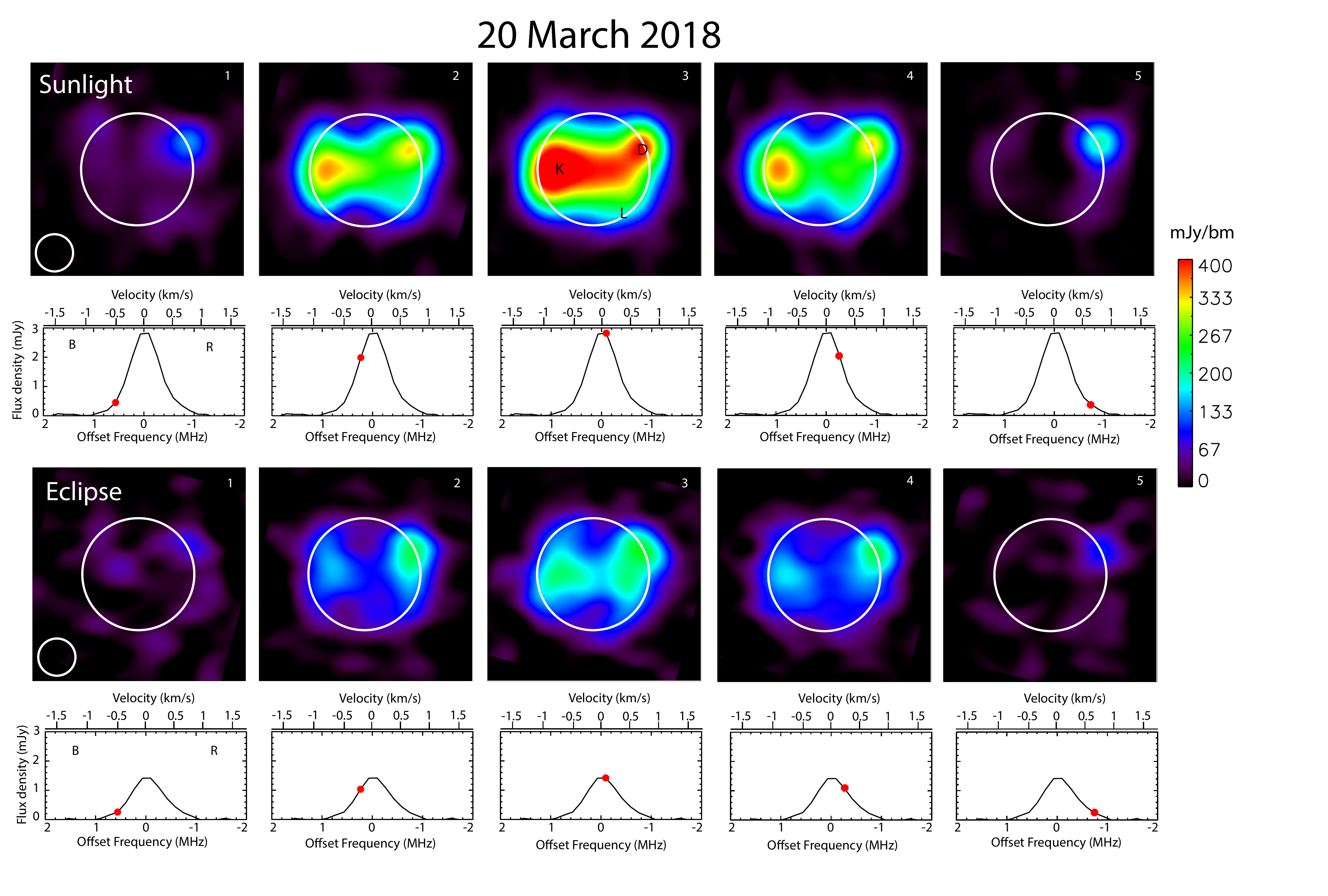}
\caption{Individual frames at a few different frequencies (or
  velocities) from our March sunlight $\rightarrow$ eclipse data for
  the combined SO$_2$ spw1 \& spw2 data. All scans 7--15 were
  averaged for the in-eclipse (Set 2; Table 2) and separately for the
  in-sunlight data. Each frame is averaged over 0.142 km/s or $\sim$~0.16
  MHz, and the line is centered on Io's frame of reference. Below each frame we show the line profile for the
  disk-integrated flux density as a function of offset frequency (from +2 to
  --2 MHz), with an
  approximate velocity scale at the top. The red dot on the line profile indicates the frequency of
  the map above. The symbols B
  and R stand for blueshift and redshift, resp., i.e., gas moving towards
  (B) or away from us (R). Note that, just due to the rotation
  of Io, the west limb (left side of
  Io) moves towards us, and the east limb away from us. The
  approximate positions of several volcanoes are indicated on frame 3
  in-sunlight (see Fig.~\ref{fig:MarchSO2} for the symbols). Io
  North is up in these frames.
An animation of this figure is available. The video shows a 14 image sequence in sunlight (left) and in eclipse (right). The duration of the video is 7 seconds.
  }
\label{fig:MarchFrames}\end{figure*}

\begin{figure*}
\includegraphics[scale=0.48]{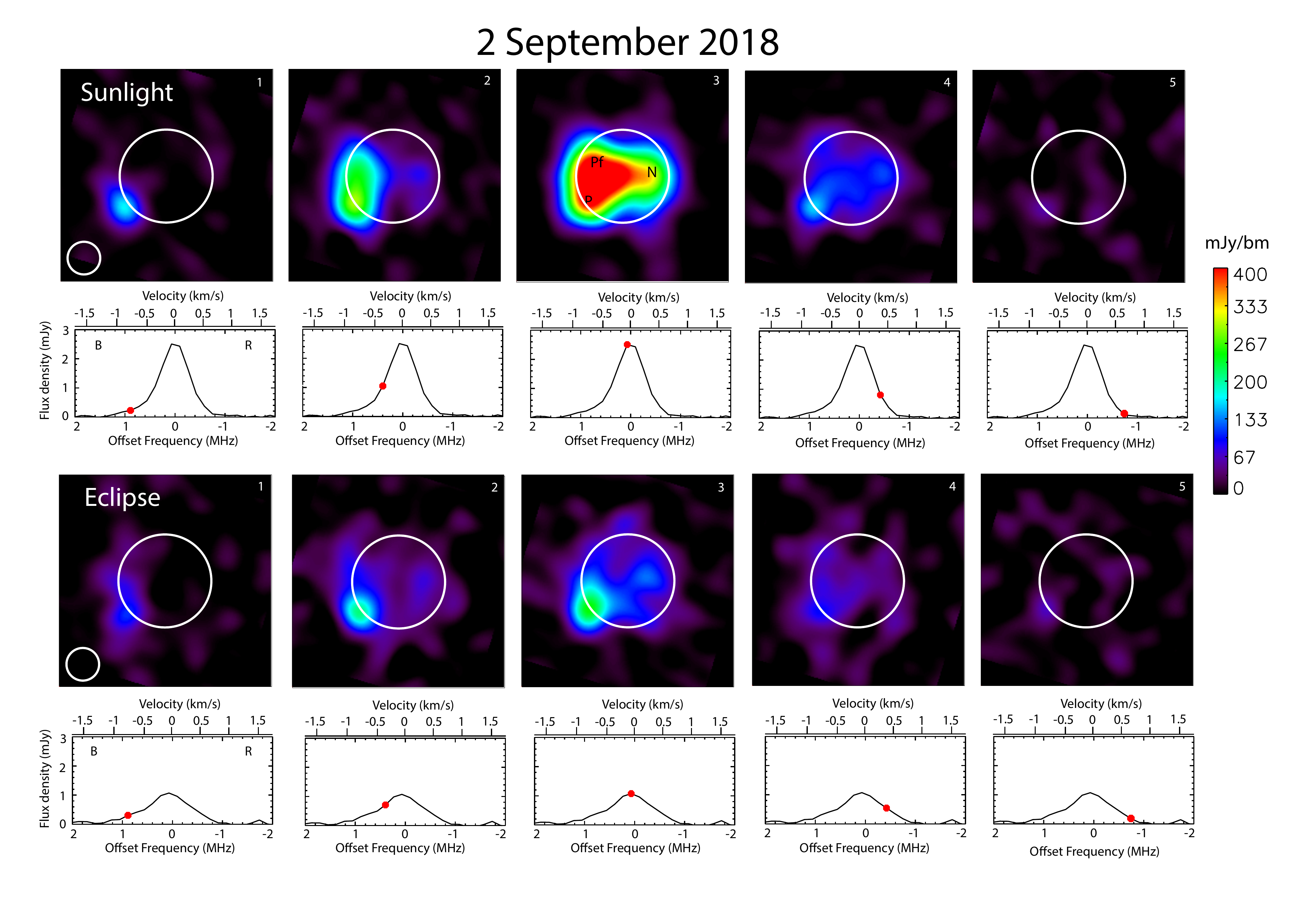}
\caption{Individual frames at a few different frequencies (or
  velocities) from our 2 September eclipse $\rightarrow$ sunlight data
  for
  the combined SO$_2$ spw1 \& spw2 data. For the eclipse data we show
  results for scan 6 only (6a$+$6b); the sunlight scans are for set 2
  (see Table 2). Each frame is averaged over 0.142 km/s or $\sim$~0.16
  MHz, and the line is centered on Io's frame of reference. Below each frame we show the line profile for the
  disk-integrated flux density as a function of offset frequency (from +2 to
  --2 MHz), with an
  approximate velocity scale at the top. The red dot indicates the frequency of
  the map above.  The symbols B
  and R stand for blueshift and redshift, resp., i.e., gas moving towards
  (B) or away from us (R). Note that, just due to the rotation
  of Io, the west limb (left side of
  Io) moves towards us, and the east limb away from us. The
  approximate positions of several volcanoes are indicated on frame 3
  in-sunlight (see Fig.~\ref{fig:Sep2SO2} for the symbols). Io
  North is up in these frames.
An animation of this figure is available. The video shows a 16 image sequence in sunlight (left) and in eclipse (right). The duration of the video is 8 seconds.
}
\label{fig:Sep2Frames}\end{figure*}
  
\begin{figure*}
\includegraphics[scale=0.48]{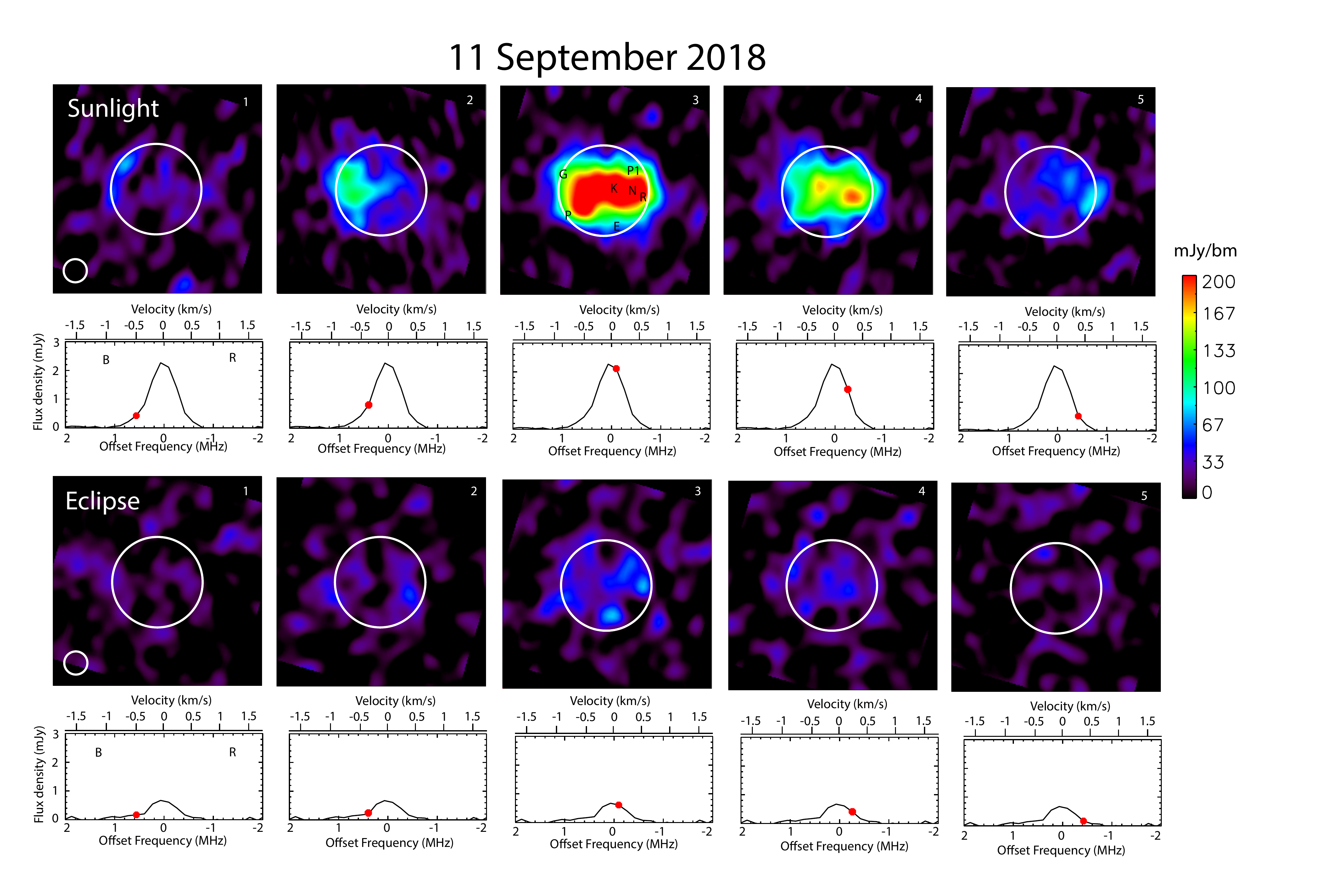}
\caption{Individual frames at a few different frequencies (or
  velocities) from our 11 September eclipse $\rightarrow$ sunlight
  data for
  the combined SO$_2$ spw1 and spw2 data. Each frame is averaged over 0.142 km/s or $\sim$~0.16
  MHz, and the line is centered on Io's frame of reference. Below each frame we show the line profile for the
  disk-integrated flux density as a function of offset frequency (from +2 to
  --2 MHz), with an
  approximate velocity scale at the top. The red dot indicates the frequency of
  the map above.  The symbols B
  and R stand for blueshift and redshift, resp., i.e., gas moving towards
  (B) or away from us (R). Note that, just due to the rotation
  of Io, the west limb (left side of
  Io) moves towards us, and the east limb away from us. The
  approximate positions of several volcanoes are indicated on frame 3
  in-sunlight (see Fig.~\ref{fig:Sep11SO2} for the symbols). Io
  North is up in these frames.
An animation of this figure is available. The video shows a 11 image sequence in sunlight (left) and in eclipse (right). The duration of the video is 6 seconds.
}
  \label{fig:Sep11Frames}\end{figure*}

\begin{figure*}
\includegraphics[scale=0.70]{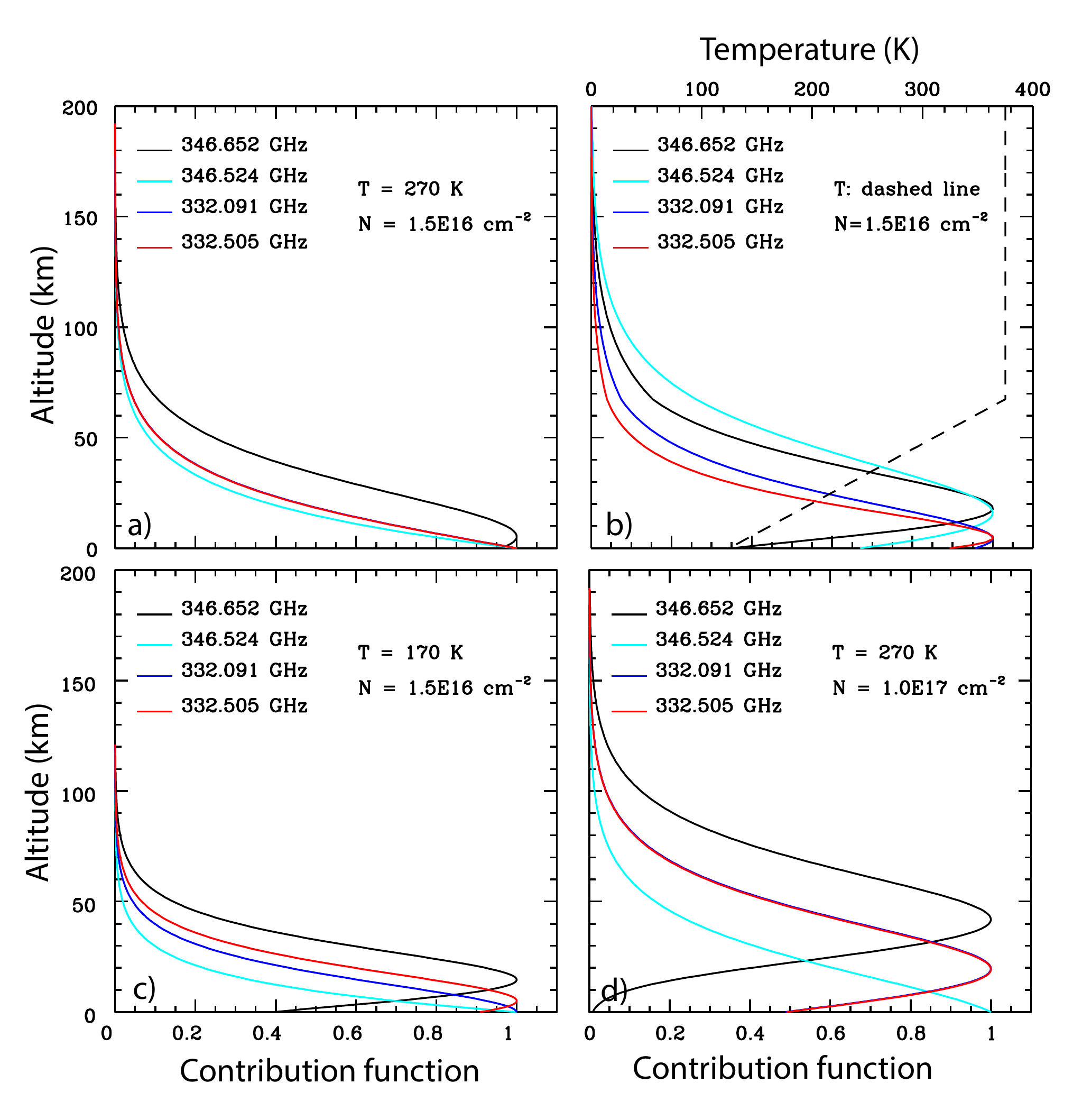}
\caption{Sample disk-averaged contribution functions for the four SO$_2$ transitions
  detected in our ALMA data, based upon our uniform, hydrostatic model atmosphere. a) Contribution functions for a column
  density ($N_t=1.5\times 10^{16}$ cm$^{-2}$) and isothermal temperature
  ($T=270$ K) that match most of our data (Sections~\ref{sec4a},
  \ref{sec4b}).
 b) A temperature profiles as indicated by the dashed line. This
 profile is inspired by profiles affected by plasma heating from
 above, such as shown by Gratiy et al. (2010). Resulting line profiles
 do {\it not} match our data. c) Contribution functions from panel a
 for a much colder isothermal atmosphere ($T=170$ K). Resulting line profiles
 do {\it not} match our data. d) Contribution functions from panel a
 for a much higher column abundance ($N_t= 10^{17}$ cm$^{-2}$), such as expected
 on the anti-jovian hemisphere. Resulting line profiles 
 do {\it not} match our data.
  }
\label{fig:contr}\end{figure*}

\begin{figure*}
\includegraphics[scale=0.65]{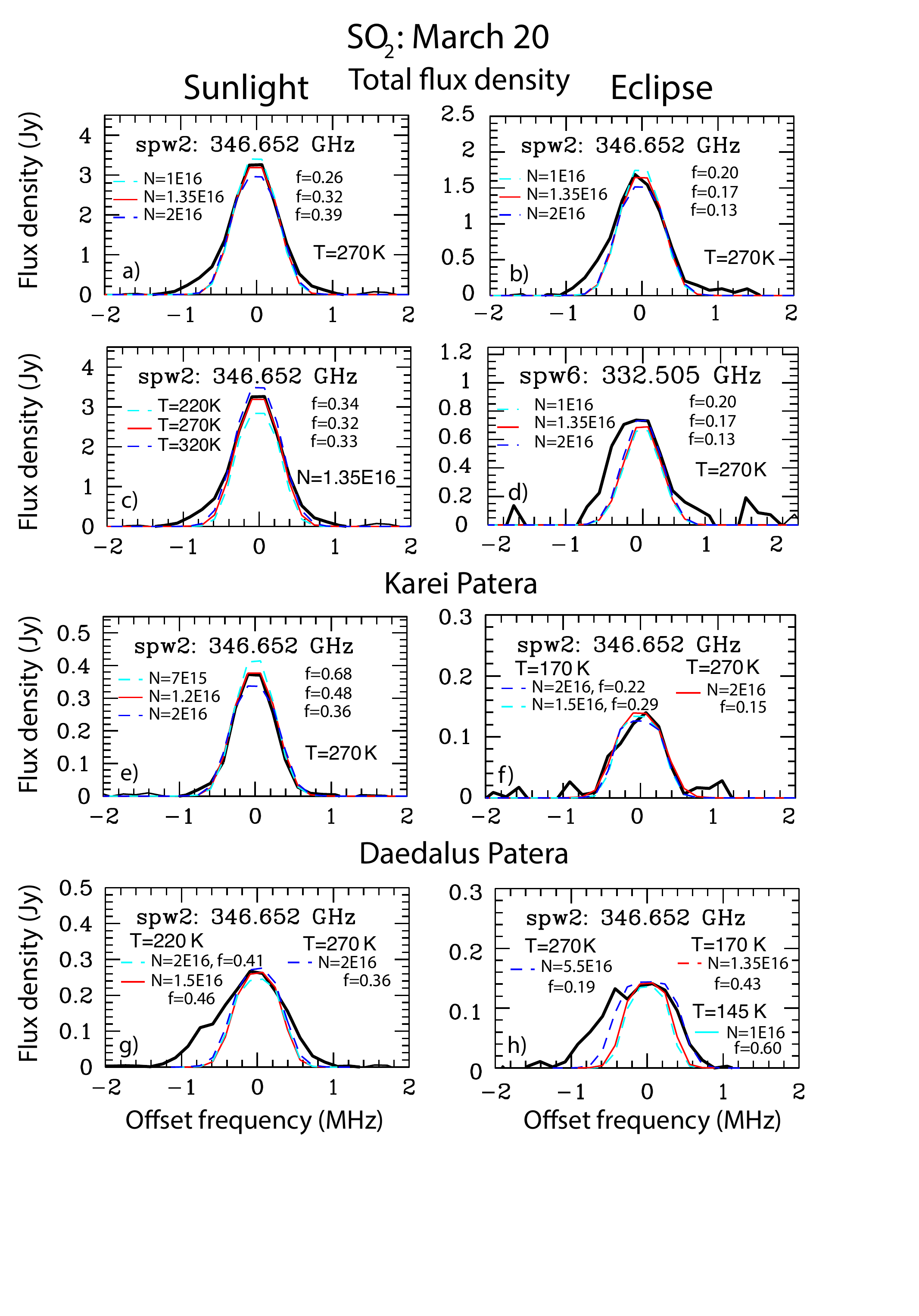}
\caption{SO$_2$ line profiles (in black) with superposed various models. The red lines
  show the best fit models. All panels show data and models at 346.652
  GHz (spw2), except for panel d.
  a) Disk-integrated flux density for Io-in-sunlight,
  with superposed the best fit ($N_t=1.35\times 10^{16}$ cm$^{-2}$)
  model at the best fit temperature $T_t=270$ K, and a fractional
  coverage $fr_t=0.32$ (in red). Several
  models are shown to provide a sense on the accuracy of the
  numbers; the fractional coverage for these models is indicated on
  the right side of the line profile. b) Disk-integrated
  flux density for Io-in-eclipse, with superposed the best
  fit ($N_t=1.35\times 10^{16}$ cm$^{-2}$), $T_t=270$ K, $fr_t=0.17$  (in red). c) Same as panel
  a to show the sensitivity on the temperature. d) Same as
  panel b, but at 332.505 GHz (spw6). e--h): data for Karei and Deadalus Paterae,
  integrated over 1 beam diameter (in black). Various hydrostatic models are
  superposed, as indicated.
  }
  \label{fig:lineprofilesM}

\end{figure*}

\begin{figure*}
\includegraphics[scale=0.65]{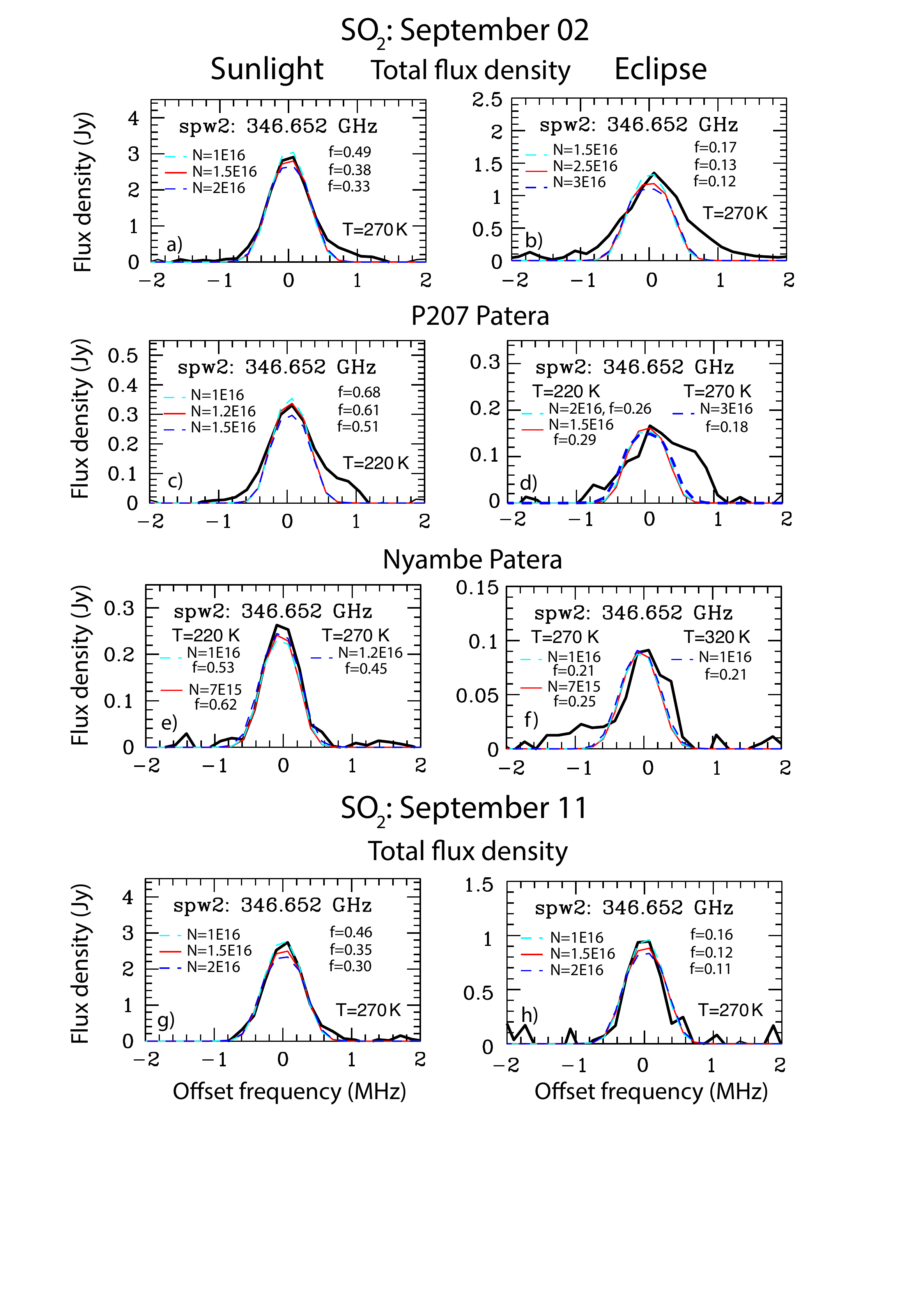}
\caption{SO$_2$ line profiles (in black) with superposed various models. The red lines
  show the best fit models. All data and models are at 346.652 GHz
  (spw2). The temperature ($T$), column density ($N$), and
  fractional coverage ($fr$) are indicated for each model. Panels
  a--f) are for 2 September, g--h) for 11 September.
  a) Disk-integrated flux density for Io-in-sunlight.
  b)  Disk-integrated flux density for Io-in-eclipse.
  c-- d) Line profiles for P207 Patera in-sunlight and in-eclipse.
  e-- f) Line profiles for Nyambe Patera in-sunlight and in-eclipse.
  g--h) Line profiles for the total flux density for 11 September
  in-sunlight and in-eclipse, respectively.
  }
\label{fig:lineprofilesS}\end{figure*}

\begin{figure*}
\includegraphics[scale=0.65]{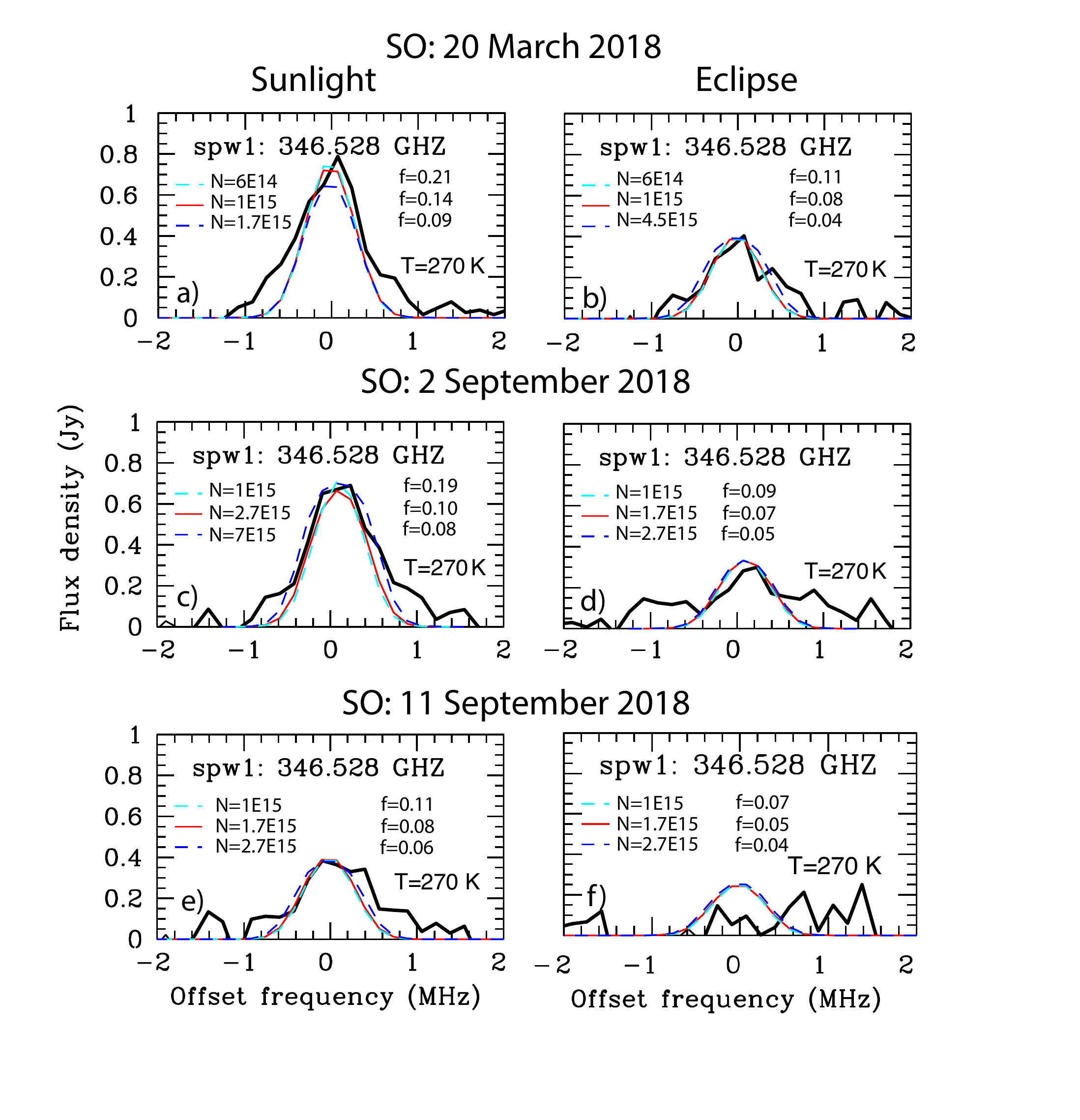}
\caption{SO line profiles (in black) with superposed various hydrostatic models. The red lines
  show the best fits; all models shown are at 346.528 GHz (spw1). The temperature ($T$), column density ($N$), and
  fractional coverage ($fr$) are indicated for each model. On the left are
  SO profiles in-sunlight; on the right in-eclipse. The dates are
  indicated above each row. Note the shape of
  the profiles, and the variations in intensity.
  }
\label{fig:lineprofilesSO}\end{figure*}

\vfill

\end{document}